\documentclass[fleqn,usenatbib]{mnras}

\usepackage{newtxtext,newtxmath}
\usepackage{multirow}
\usepackage{hyperref}
\usepackage{subfigure}

\usepackage[dvipsnames]{xcolor} 
\usepackage[normalem]{ulem} 
\definecolor{darkgreen}{rgb}{0.0,0.75,0.0}
\definecolor{darkblue}{rgb}{0.0,0.0,0.5}
\definecolor{orange}{rgb}{1.0,0.65,0.0}
\definecolor{grey}{gray}{0.7}

\usepackage[T1]{fontenc}

\DeclareRobustCommand{\VAN}[3]{#2}
\let\VANthebibliography\thebibliography
\def\thebibliography{\DeclareRobustCommand{\VAN}[3]{##3}\VANthebibliography}

\usepackage{graphicx}	
\usepackage{amsmath}	

\title{First IFU observations of two GRB host galaxies at cosmic noon with JWST/NIRSpec}
\author[B. Top\c{c}u et al.]{B. Top\c{c}u$^{1}$\thanks{E-mail: bt434@bath.ac.uk},
P. Schady$^{1}$, S. Wuyts$^{1}$, A. Inkenhaag$^{1}$, M. Arabsalmani$^{2,3}$, H.-W. Chen$^{4}$, L. Christensen$^{5,6}$,
\newauthor
V. D'Elia$^{7}$, J. P. U. Fynbo$^{5,6}$, K. E. Heintz$^{5,6}$, P. Jakobsson$^{8}$, T. Laskar$^{9,10}$, A. Levan$^{10}$, G. Pugliese$^{11}$, 
\newauthor
A. Rossi$^{12}$, R. L. C. Starling$^{13}$, N. R. Tanvir$^{13}$, P. Wiseman$^{14}$, R.M. Yates$^{15}$
\\
$^{1}$ Departmant of Physics, University of Bath, Claverton Down, Bath BA2 7AY, UK \\
$^{2}$ Excellence Cluster ORIGINS, Boltzmannstraße 2, 85748 Garching, Germany \\
$^{3}$ Ludwig-Maximilians-Universität, Schellingstraße 4, 80799 München, Germany \\
$^{4}$ Department of Astronomy \& Astrophysics, The University of Chicago, 5640 S Ellis Ave., Chicago, IL 60637, USA \\
$^{5}$ Niels Bohr Institute, University of Copenhagen, Jagtvej 128, DK-2200 N, Copenhagen, Denmark \\
$^{6}$ Cosmic Dawn Center (DAWN), Denmark\\
$^{7}$ Space Science Data Center (SSDC) - Agenzia Spaziale Italiana (ASI), I-00133 Roma, Italy \\
$^{8}$ Centre for Astrophysics and Cosmology, Science Institute, University of Iceland, Dunhagi 5, 107 Reykjavik, Iceland \\
$^{9}$ Department of Physics \& Astronomy, University of Utah, Salt Lake City, UT 84112, USA \\
$^{10}$ Department of Astrophysics/IMAPP, Radboud University, 6525 AJ Nĳmegen, The Netherlands \\
$^{11}$ Anton Pannekoek Institute for Astronomy, University of Amsterdam, P.O. Box 94249, 1090GE Amsterdam, The Netherlands \\
$^{12}$ INAF – Osservatorio di Astrofisica e Scienza dello Spazio, Via Piero Gobetti 93/3, 40129 Bologna, Italy \\
$^{13}$ School of Physics and Astronomy, University of Leicester, University Road, Leicester, LE1 7RH, UK \\
$^{14}$ School of Physics and Astronomy, University of Southampton, Southampton, SO17 1BJ, UK \\
$^{15}$ Centre for Astrophysics Research, University of Hertfordshire, Hatfield, AL10 9AB, UK 
}

\date{Accepted XXX. Received YYY; in original form ZZZ}

\pubyear{2014}

\begin{document}
\label{firstpage}
\pagerange{\pageref{firstpage}--\pageref{lastpage}}
\maketitle

\begin{abstract}
Long gamma-ray bursts (GRBs) serve as powerful probes of distant galaxies. Their luminous afterglow pinpoints galaxies independent of luminosity, in contrast to most flux-limited surveys. Nevertheless, GRB-selected galaxy samples are not free from bias, instead tracing the conditions favoured by the progenitor stars. Characterising the galaxy populations traced by GRBs is therefore important both to effectively use GRBs as probes as well as to place stronger constraints on the progenitor stars capable of forming long GRBs. Spatially-resolved spectroscopic observations with integral field units (IFUs) provide valuable insights into the interstellar medium and stellar populations of GRB host galaxies. In this paper we present results of the first two GRB host galaxies observed with the {\em JWST}/NIRSpec IFU with a spatial resolution of $\sim 1.6$~kpc; the hosts of GRB 150403A and GRB 050820A at redshifts $z\sim 2.06$ and $z\sim 2.61$, respectively. The data reveal two complex galaxy environments made up of two or more star forming galaxies that are likely interacting given their small spatial separation ($<20$~kpc) and line of sight velocity offsets ($<100$~km/s). The measured gas-phase metallicity, star formation rates (SFRs), and key diagnostic line ratios for each of the detected galaxies are overall consistent with the properties of other star forming galaxies and GRB hosts at $z>2$. However, differences in the SFR and metallicities of the interacting galaxies highlight the importance of spatially resolved observations in order to accurately characterise the galaxy properties traced by GRBs.
\end{abstract}




\section{Introduction}
Gamma-ray bursts (GRBs) are extremely luminous, short-lived outbursts of gamma rays that reach energies of $10^{49}$ - $10^{54}$ ergs \citep[e.g,][]{berger2005_GRBenergy, zhang2009_mostenergeticGRB}, and were first detected serendipitously in the late 1960s \citep{klebesadel+1973_firstGRBobs}. Observations reveal a bimodal distribution of GRB durations, leading to the classification of GRBs into two populations: long duration GRBs with prompt $\gamma$-ray emission lasting longer than 2 seconds, and short duration GRBs with emission shorter than 2 seconds. It is now widely accepted that long GRBs are associated with the core collapse of a massive star \citep{woosley2006_SNeGRB_connection} whereas short-GRBs are formed from compact binary star mergers \citep{eichler1989_GRBmerger_model, narayan1992_GRB_NSmerger, abbott+2017_GRBandGW} \citep[although see e.g.][]{rastinejad2022_LGRBmerger,troja2022_LGRBmerger,levan2024_LGRBmerger,yang2024_LGRBmerger}. Long GRBs are therefore powerful tracers of star formation since they select star-forming galaxies independent of galaxy luminosity \citep[e.g.,][]{berger2003_GRBhostSFR, le2003_GRBhostSFR, Christensen_2004_GRBhostSFR, kistler2009_bias_cosmicSFR, savaglio2009_GRBhostSFR}, and can be detected out to very high redshifts \citep[e.g,][]{greiner2009_highz6GRB, salvaterra2009_highz8GRB, tanvir2009_highz8GRB}. The successful, long GRB `collapsar' model \citep{woosley1993_GRB+collapsarmodel} nevertheless predicts a cap on the metallicity of the progenitor star ($<0.3Z_\odot$), which would introduce selection effects in GRB-selected galaxy samples. Characterising the properties of long GRB host galaxies is therefore not only important to test key predictions from GRB progenitor models, but it is also an essential step in order to understand the galaxy properties that are traced by long GRBs. 

With a handful of exceptions \citep[e.g.,][]{rossi2014_LGRB_dimhost, levan2023noSN_GRB}, long GRBs are found in low luminosity, star-forming, low mass galaxies \citep[e.g.,][]{le2003_GRBhostSFR,fruchter2006_GRB_massive, kelly+2008longGRBlocations, savaglio2009_GRBhostSFR,svensson2010grbvsccsne_host,li+2011_SNrates,perley2016_GRB_SFR}, with a preference for lower metallicity environments \citep[e.g.,][]{le2003_GRBhostSFR, tanvir2004_bias/selection, savaglio2009_GRBhostSFR, levesque2010a_GRBhostgalaxies, levesque2010b_highZ_host, graham2013_GRBsurvey_metal, kruhler2015grb,graham_2017metallicitycap}, in agreement with the expectations of the collapsar model. The metallicity drop off, however, does not appear as a sharp cut-off, and a notable fraction of long GRBs have been found to reside within more massive, chemically enriched, and dusty galaxies \citep[e.g.,][]{fruchter2006_GRB_massive,kruhler2011_GRB_SFR, rossi2012_darkGRB_SFR, perley2016_GRB_SFR, tanga2018_museGRB}. Any metallicity bias should also become less important at higher redshift, when the Universe was less chemically enriched. However, due to the increasing difficulty in obtaining good quality spectra to measure the metallicity of GRB host galaxies at $z>2$ \citep[e.g.,][]{kruhler2015grb}, it remains unclear at what redshift long GRBs are no longer biased tracers of star formation. 

In addition to long GRBs showing a preference for low metallicity environments, high spatial resolution \textit{Hubble Space Telescope} ({\rm HST}) observations reveal a strong tendency for long GRB host galaxies to exhibit irregular morphologies \citep{fruchter2006_GRB_massive}, with a significant fraction appearing to be interacting or merging systems \citep{conselice_2005morphologies,wainwright_2007morphologies}. Complementary to HST observations, \ion{H}{i} 21~cm emission line studies have identified GRB hosts with clear signs of interactions \citep[e.g.,][]{arabsalmani2015_980425interact, arabsalmani2019_980425interact, arabsalmani2022_171205Ainteract}. Additionally, the detection of multiple absorbers along the line of sight in optical afterglow spectra at close velocity offsets from the GRB \citep[e.g.,][]{savaglio2002_nearabsorbers, klose2004absorbers, ferrero2009absorbers_IFS, page2009_absrobers, savaglio2012supersolar_interacting} provides further indication that interaction may be a common feature of long GRB host galaxies, especially at $z>2$. A possible reason for this is that galaxy interactions may trigger a starburst episode, leading to a rise in the formation of massive stars and thus in the likelihood of a long GRB event \citep{somerville2001_interaction+starburst, lopez2009interaction_SF, teyssier2010_int+starburst+sim}. Such merger interactions can also induce gas instabilities that funnel low-metallicity gas into central regions of galaxies, causing a starburst in the less enriched gas compared to the overall host galaxy ISM. This combination of elevated star formation rates and low metallicities may create particularly favourable conditions for long GRB production.

With integral field unit (IFU) instruments such as the Visible Multi Object Spectrograph \citep[VIMOS;][]{lefevre2003_VIMOS} and the Multi Unit Spectroscopic Explorer \citep[MUSE;][]{bacon2010_MUSE}, it has been possible to obtain spatially resolved spectroscopic observations of GRB host galaxies, where both galaxy morphology and spectral properties can be studied \citep[e.g.,][]{christensen2008_vimosGRB,izzo2017_museGRB,kruhler2017_museGRB,tanga2018_museGRB,thoene_2024GRB171205Ahost}. The combined sensitivity, large field of view and high spatial resolution (0\farcs 2 in wide field mode) of MUSE has made it possible, in some cases, to study the stellar population at the GRB site, revealing metal poor \ion{H}{ii} regions, with ages $<10$ Myr, implying massive long GRB progenitors ($>20$ M$_{\odot}$), in line with expectations of the collapsar model \citep[although see GRB 111005A;][]{tanga2018_museGRB}. While valuable for studying nearby GRB host galaxies, the spectral coverage of MUSE ($0.465 - 0.93\mu$m) limits the access to key diagnostic nebular lines for more distant galaxies, with [\ion{O}{iii}]$\lambda5007$ and  H$\alpha$ no longer being accessible at $z \gtrsim 1.0$.

The unprecedented sensitivity and spectral range of the \textit{James Webb Space Telescope} (\textit{JWST}) has now significantly extended the redshift out to which sensitive and spatially resolved spectroscopic observations are available with the \textit{Near-Infrared Spectrograph} \citep[NIRSpec;][]{jakobsen2022_NIRSpec1, ferruit2022_NIRSpec2, boker2022_NIRSpec3}. The operational wavelength range of NIRSpec is $0.6 - 5.3 \mu$m, providing access to strong optical line emission including H$\alpha$ out to $z \sim 7$.

Here we present the results from {\em JWST} NIRSpec integral field spectrograph \citep[IFS;][]{jakobsen2022_NIRSpec1} observations of the host galaxies of the two long GRBs, GRB 050820A and GRB 150403A at redshifts z $\sim 2.65$ and $\sim 2.06$ respectively. These data were included in \cite{schady2024comparing} as part of a larger spectroscopic sample of GRB host galaxies at $z\gtrsim 2$ (mostly spatially unresolved) that focused on the relation between GRB afterglow absorption and host galaxy emission line metallicities. In this work we present a more detailed analysis of the spatially resolved emission line properties of the two targets, focusing on the dynamical properties and presenting a comparison with previously studied samples of long GRB host galaxies and high redshift star-forming galaxies. In section~\ref{sec:2} we provide details on the NIRSpec observations and describe our data analysis in section~\ref{sec:3}. In section~\ref{sec:result} we present our results on the kinematics and emission line properties of the two galaxies, and in section~\ref{sec:discussion} we discuss the implications of our results and compare the two galaxies to a larger sample of $z>2$ long GRB host galaxies and to the general star forming galaxy population. Finally, we provide our conclusions in section~\ref{sec:conclusion}. All uncertainties are given as $1\sigma$ unless otherwise stated and we assume a standard Lambda cold dark matter ($\Lambda$CDM) cosmology with $\Omega_{M}$ = 0.31, $\Omega_{\Lambda}$ = 0.69, and H$_{0}$ = 67.8 km $\mathrm{s}^{-1} \mathrm{Mpc}^{-1}$ \citep{2016A&A...594A..13P}.

\section{Observations}\label{sec:2}
The host galaxies of GRB 050820A and GRB 150403A were both part of a larger GRB host galaxy sample observed during a cycle-1 {\em JWST}/NIRSpec program \cite[ID 2344;][]{schady2024comparing}. They were the only two galaxies in the sample to be observed using the IFS, prompted by the extended emission seen in pre-existing imaging of GRB 050820A and GRB 150403A with HST \citep{chen2012_GRB050820A_HST} and the GRB Optical and Near-infrared Detector \citep[GROND;][]{greiner2008Grond}, respectively. The {\em JWST} NIRSpec observations provide spatially resolved spectroscopic information on pixel scales $0\farcs1$ over a $3'' \times$ $3''$ field of view. The IFS point spread function full width half maximum is around 2 pixels wide, corresponding to a physical spatial resolution of $\sim 1.6$~kpc for both host galaxies. Observations were taken using the G140M/F100LP and G235M/F170LP grating/filter combinations corresponding to a wavelength coverage of 1.0 -- 1.9 $\mu$m and 1.7 -- 3.2 $\mu$m respectively, which at the redshift of the two host galaxies, covered all strong emission lines from [\ion{O}{ii}] $\lambda\lambda$3726,3729 to [\ion{S}{ii}]  $\lambda\lambda$6717,6731. The spectral resolution ranges from R$ \sim $500 to R$ \sim $1500, corresponding to a velocity width resolution of $\sigma\sim 85-250$~km~s$^{-1}$.

The reduced and calibrated 3D spectra were downloaded from the Mikulski Archive for Space Telescopes (MAST) Data Discovery Portal \footnote{\url{https://mast.stsci.edu/portal/Mashup/Clients/Mast/Portal.html}}. The data were reduced with version 11.17.2 of the CRDS file selection software, using context jwst\textunderscore1140.pmap, and were calibrated with version 1.11.4 of the calibration software. The reader is referred to \cite{schady2024comparing} for further details regarding the observations of the targets.

\section{Data Analysis}\label{sec:3}
\subsection{Gaussian line fitting}
For both targets, emission from H$\alpha$, H$\beta$, [\ion{O}{ii}]$\lambda\lambda3726, 3729$ and [\ion{O}{iii]}$\lambda\lambda4959,5007$ are detected, but there is no significant emission at the position of [\ion{N}{ii}]$\lambda\lambda6549, 6584$ or [\ion{S}{ii}]$\lambda\lambda6717,6731$. The galaxy continuum is not detected in either spectra.
Line fluxes for each spaxel were measured from Gaussian fits to the emission lines. For greater constraint, the fits were performed simultaneously for H$\beta$ and [\ion{O}{iii}]$\lambda\lambda4959,5007$, with the peak positions of all three lines tied to a consistent redshift and keeping the velocity widths also tied. Additionally, the flux ratio of the [\ion{O}{iii}]$\lambda\lambda4959,5007$ doublet was fixed at 1:3, as per \cite{osterbrock1989active}. The velocity width and peak position of all other fitted lines were then fixed to be consistent with the best-fit values from the H$\beta$ and [\ion{O}{iii}]$\lambda\lambda4959,5007$ line fits. Although the [\ion{O}{ii}]$\lambda\lambda3726, 3729$ doublet was not resolved, a double Gaussian was fit to the line, and the total flux was taken as the sum of the two profiles. In those cases where the velocity width of $\lambda\lambda4959,5007$ was unresolved, the fitted full width half maximum (FWHM) was left as a free parameter in subsequent line fits. Although [\ion{N}{ii}]$\lambda6584$ was undetected in the observations in both galaxies, a flux upper limit was determined from simultaneous fits to H$\alpha$ and [\ion{N}{ii}]$\lambda6584$, with the peak position and velocity width tied as described above. The flux upper limit was then determined from the best fit amplitude and $1\sigma$ uncertainty at the position of [\ion{N}{ii}]$\lambda6584$.

All line fluxes were corrected for Milky Way dust extinction using the \cite{schlafly2011_MWebv} $E(B-V)$ reddening maps, which gave $E(B-V)=0.038 \pm 0.001$ for GRB~050820A and $E(B-V)=0.047 \pm 0.001$ for GRB~150403A. The extinction correction was performed assuming a \cite{cardelli_ebv_law} dust extinction curve with an average total-to-selective dust extinction value $R_V = 3.08$.

The host galaxy dust reddening was calculated from the H$\alpha$/H$\beta$ Balmer decrement assuming an intrinsic ratio H$\alpha$/H$\beta$=2.86 \citep{osterbrock1989active} and using the average Small Magellanic Cloud (SMC) dust law from \citet{pei1992_dustlaws} with $R_V=2.93$. In some cases the measured host galaxy $E(B-V)$ was less than zero, which is not physical, and the $E(B-V)$ was therefore set to zero when applying dust corrections. 

\subsection{Emission line fluxes}\label{sec:3.2}
\begin{figure}
    \centering
    \includegraphics[width = \columnwidth]{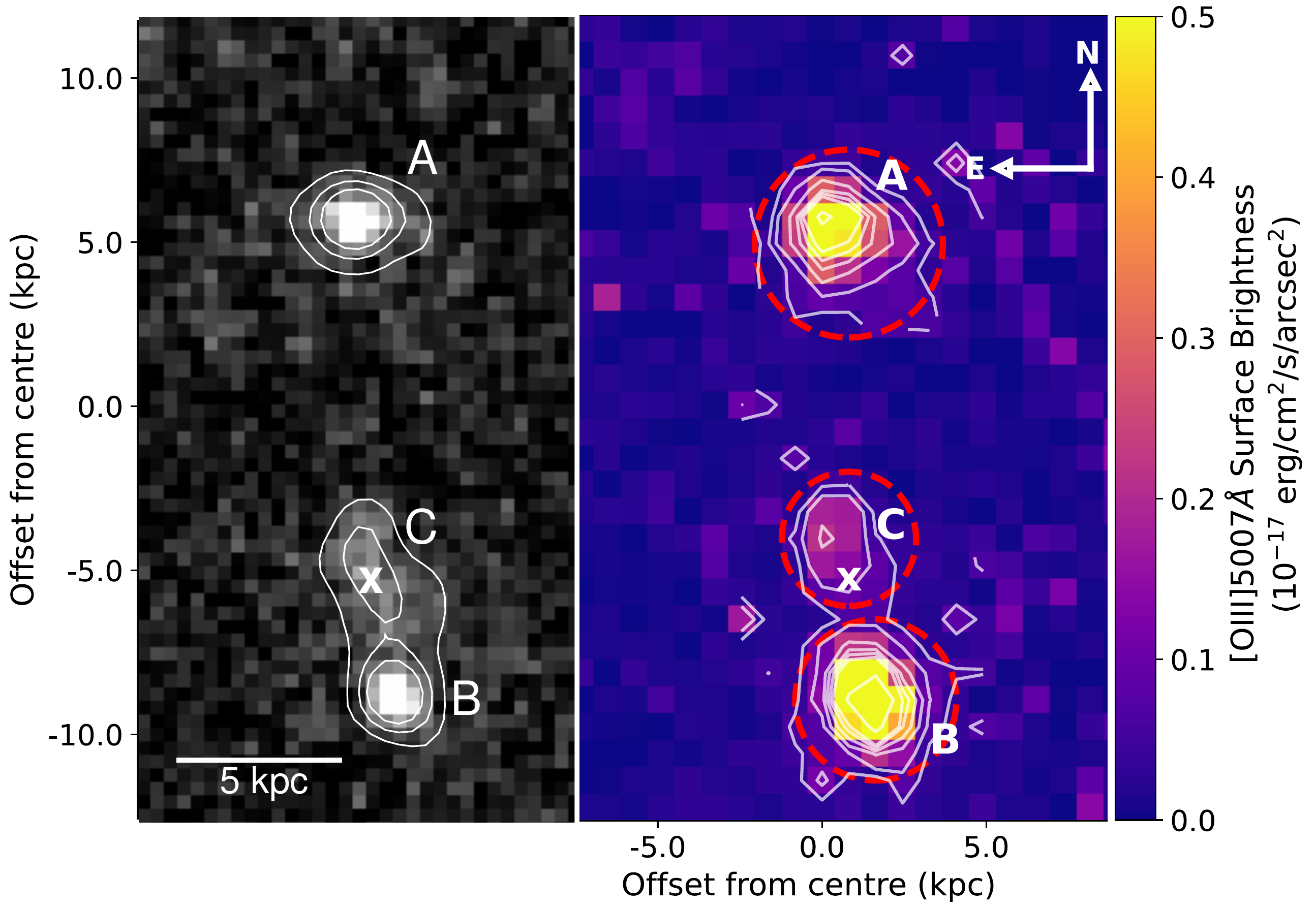}
    \caption{Image of the field of the host galaxy of GRB~050820A showing continuum emission at {\em F775W} observed with the {\em HST}/Advanced Camera for Surveys (ACS) (PID 10551; PI: S. Kulkarni) (\textit{left panel}) and [\ion{O}{iii}]$\lambda 5007$ flux emission from our {\em JWST}/NIRspec data (\textit{right panel}). In both images we label the star forming components A and B identified in \citet{chen2012_GRB050820A_HST}, as well as the fainter component C. The position of the GRB detected in early-epoch {\em HST} observations \citep{chen2012_GRB050820A_HST} is indicated with an `X'. The pixel scale of the {\em HST} and {\em JWST} images is  0\farcs05 and 0\farcs1, respectively, and the uncertainty on the GRB position is within a couple of ACS pixels and a single NIRSpec pixel. For clarity, contours tracing the the {\em F775W} continuum at flux levels 0.6, 0.65 and 0.7 $\times 10^{-20}$~erg~cm$^{-2}$~s$^{-1}$, and the [\ion{O}{iii}]$\lambda5007$ surface brightness levels 0.05, 0.1, 0.2, 0.3, 0.4, 0.5 and 1.0 $\times 10^{-17}$~erg~cm$^{-2}$~s$^{-1}$~arcsec$^{-2}$ are shown in white. Three dashed red circles in the right-hand panel indicate the sizes of the apertures used in our stacking analysis (see section \ref{sec:3.2}). Both images are oriented with North up and East to the left, and the physical scale is indicated in both panels.}
    \label{fig:grb050820_flux}
\end{figure}

\begin{figure}
    \centering
    \includegraphics[width = \columnwidth]{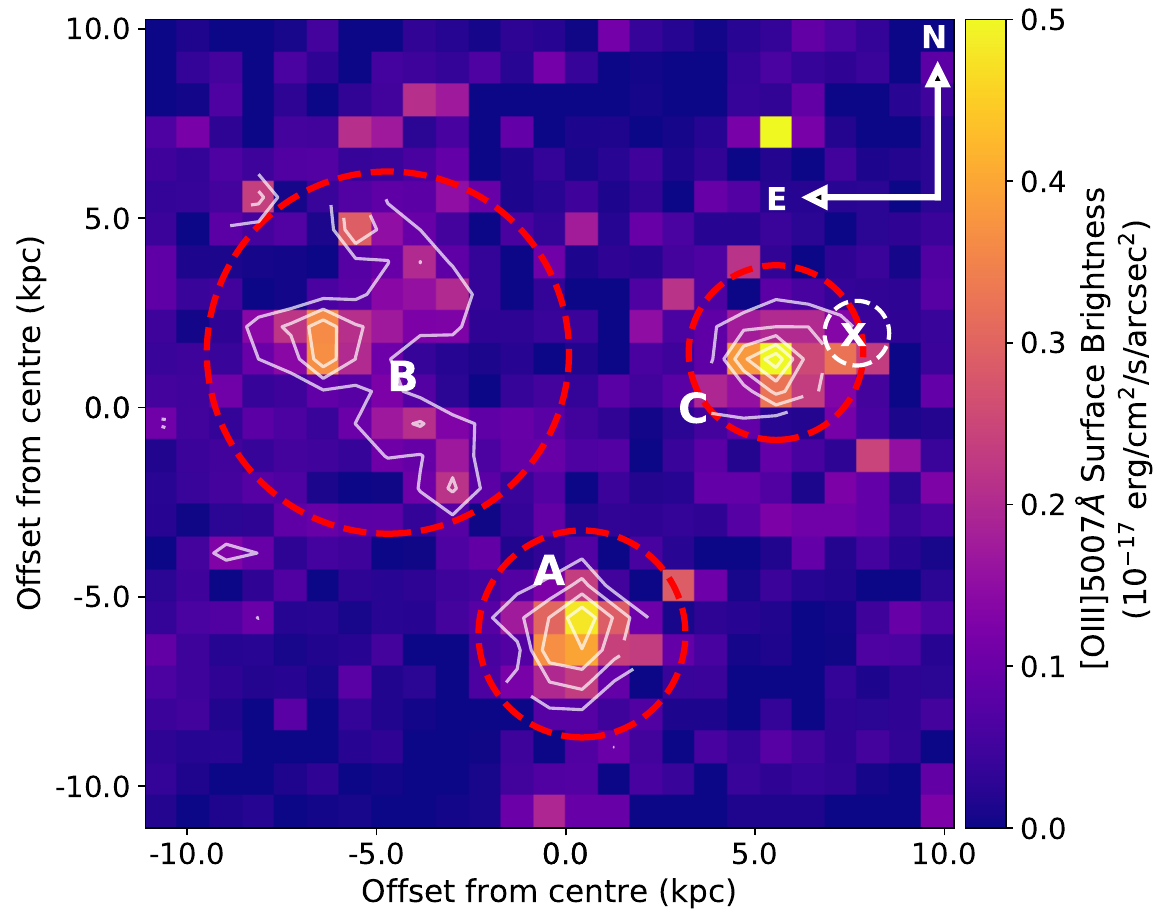}
    \caption{[\ion{O}{iii}]$\lambda5007$ surface brightness map of the host galaxy of GRB 150403A. Three star-forming components are labelled as A, B and C. The position of the GRB afterglow is just west of component C indicated with an 'x'. The $1\sigma$ uncertainty on the GRB position is indicated by the dashed circle. Some bright emission is observed to the north of C which is due to noise. Similarly, contours tracing the flux strength from 0.1 to 0.5 $\times 10^{-17}$ erg~cm$^{-2}$~s$^{-1}$~arcsec$^{-2}$ are shown in white and the extraction regions used for stacking (see section \ref{sec:3.2}) are shown with dashed red circles. The pixel scale of the image is 0\farcs1.}
    \label{fig:grb150403_o3flux}
\end{figure}

Surface brightness maps were created for all detected lines (H$\alpha$, H$\beta$, [\ion{O}{ii}]$\lambda\lambda3726,3729$ and [\ion{O}{iii}]$\lambda\lambda4959,5007$). As an example, in Fig.~\ref{fig:grb050820_flux} and Fig.~\ref{fig:grb150403_o3flux} we show the [\ion{O}{iii}]$\lambda5007$ surface brightness map for the two targets. In Fig.~\ref{fig:grb050820_flux}, we also include an {\em HST} image of GRB~050820A (left panel), showing continuum emission in the {\em F775W} filter alongside our NIRSpec [\ion{O}{iii}]$\lambda5007$ surface brightness map. The extended nature of the emission regions is apparent in both the {\em HST} image and the [\ion{O}{iii}]$\lambda5007$ surface brightness map.

Note that the surface brightness maps represent a pixel-by-pixel analysis of the flux based on Gaussian fits to the corresponding line in that spaxel, whereas in \citet{schady2024comparing} the surface brightness maps were produced by collapsing a slice of the IFU data cube centred on the emission line of interest. As was done in \cite{schady2024comparing}, we highlight three distinct emission regions in the two galaxies and label these `A', `B' and `C'. In \cite{schady2024comparing} an additional emission region was seen to the left of component C in the [\ion{O}{iii}]$\lambda5007$ map of the host galaxy of GRB 050820A, but this is no longer present with the improved NIRSpec data reduction used here, indicating that the feature was due to noise. Similar noise characteristics are observed to the north of component C in the surface brightness map of GRB 150403A in Fig.~\ref{fig:grb150403_o3flux}. These features are not present in other line maps and their spectral features resemble narrow spikes rather than Gaussian profiles. To help distinguish between regions of true line emission and noise, we show contours (white lines) in Figs.~\ref{fig:grb050820_flux} and \ref{fig:grb150403_o3flux} that trace the line flux strength. The flux map host galaxy dust corrections were applied to each of the three components separately, using the relevant E(B-V) measured from the corresponding Balmer decrement for each component, as reported in Table \ref{tab:Metal_SFR}. A component-averaged dust correction was applied rather than computing the Balmer decrement on each individual spaxels in order to increase the signal to noise ratio (SNR) of the measured E(B-V), which was too low to construct an E(B-V) pixelated map for either observation.


In addition to the pixel-by-pixel analysis, we also measured the integrated fluxes for the individual star-forming components identified above within both targets and labelled in Figs.~\ref{fig:grb050820_flux} and \ref{fig:grb150403_o3flux}. To do this we stacked the spectra within a circular aperture centred on the region of interest and with a radius sufficiently large to ensure that the entire emission region was encompassed within the aperture, see dashed red circles in Figs.~\ref{fig:grb050820_flux} and \ref{fig:grb150403_o3flux}. The apertures used had radii ranging from 0.25\arcsec to 0.6\arcsec. The stacking was performed by summing the spectra from individual spaxels within the aperture and propagating their respective uncertainties to obtain a single stacked spectrum for each component. The line fluxes in the stacked spectra were then measured using Gaussian fits and we applied Milky Way and host galaxy dust corrections as previously described. In Fig.~\ref{fig:1dspec_05} and~\ref{fig:1dspec_15} we show the Gaussian fits applied to the stacked spectra of the three components for both GRB 050820A and GRB 150403A.

We note that for the host galaxy of GRB 150403A, component B has two extended structures, and thus approximating this component by using a circular region results in a notable amount of background being included in the source aperture. To investigate the effect of the background noise on our measured line fluxes, we divided component B into three separate regions corresponding to the central circular emitting region and the two extended regions to the North-West and South-West. We then summed the line fluxes measured within each of these smaller regions. We found no significant difference in the measured line flux SNR between the two methods, and we therefore use the results from the single aperture method for further analysis. The Milky Way extinction corrected fluxes are reported in Table~\ref{tab:Host_fluxes} for each component along with the redshifts corresponding to the best-fit peak wavelength fitted to the [\ion{O}{iii}]$\lambda 5007$ line. The uncertainties on our redshift measurements include the NIRSpec wavelength calibration uncertainty, which is on the order of 0.8\AA \footnote{\url{https://jwst-docs.stsci.edu/depreciated-jdox-articles/jwst-data-calibration-considerations/jwst-data-absolute-wavelength-calibration\#gsc.tab=0}}. 

The fluxes we measure are generally in good agreement with those reported in \cite{schady2024comparing} within 1$\sigma$, and larger differences can be explained by differences in the source extraction regions used for each component. The measured emission line redshifts are consistent with the absorption line redshifts of 2.615 \citep{prochaska2007_GRB050820absredsft, ledoux2009_absredshiftGRB05} and 2.057 \citep{selsing2019_abs_redshift} measured from the afterglow spectra of GRB 050820A and GRB 150403A, respectively.

In Table~\ref{tab:Metal_SFR} we give the measured host galaxy $E(B-V)$ values for each component 
along with the measured intrinsic velocity dispersions, $\sigma$. We note that the measured $\sigma$ values correspond to FWHM comparable to the NIRSpec line spread function (LSF) and the wavelength pixel scale. To quantify the impact of spectral resolution on our best-fit line widths and peak wavelengths, we generated a Gaussian model on an oversampled wavelength grid, and then binned this to the wavelength scale of our observed spectra. We then re-fitted our spectra with this discretely-sampled Gaussian model. Comparing the two cases, we found that the newly fitted line widths were 1.4 to 2.1 times narrower than the line widths measured from an unbinned Gaussian model, although the best-fit peak wavelengths were unchanged within uncertainties. Consequently, we refrain from performing any analysis using the measured velocity widths, and merely report for completeness the (generally small or unresolved) velocity dispersions from the binned Gaussian model fits in Table~\ref{tab:Metal_SFR}. We considered any extra line broadening to be intrinsic only if it accounted for more than 10\% of the observed FWHM; otherwise, we report the sigma dispersion as an upper limit. 

\section{Results}\label{sec:result}

\subsection{Kinematics}
\begin{figure*}
    \centering
    \begin{subfigure}
        \centering
        \includegraphics[width = \columnwidth]{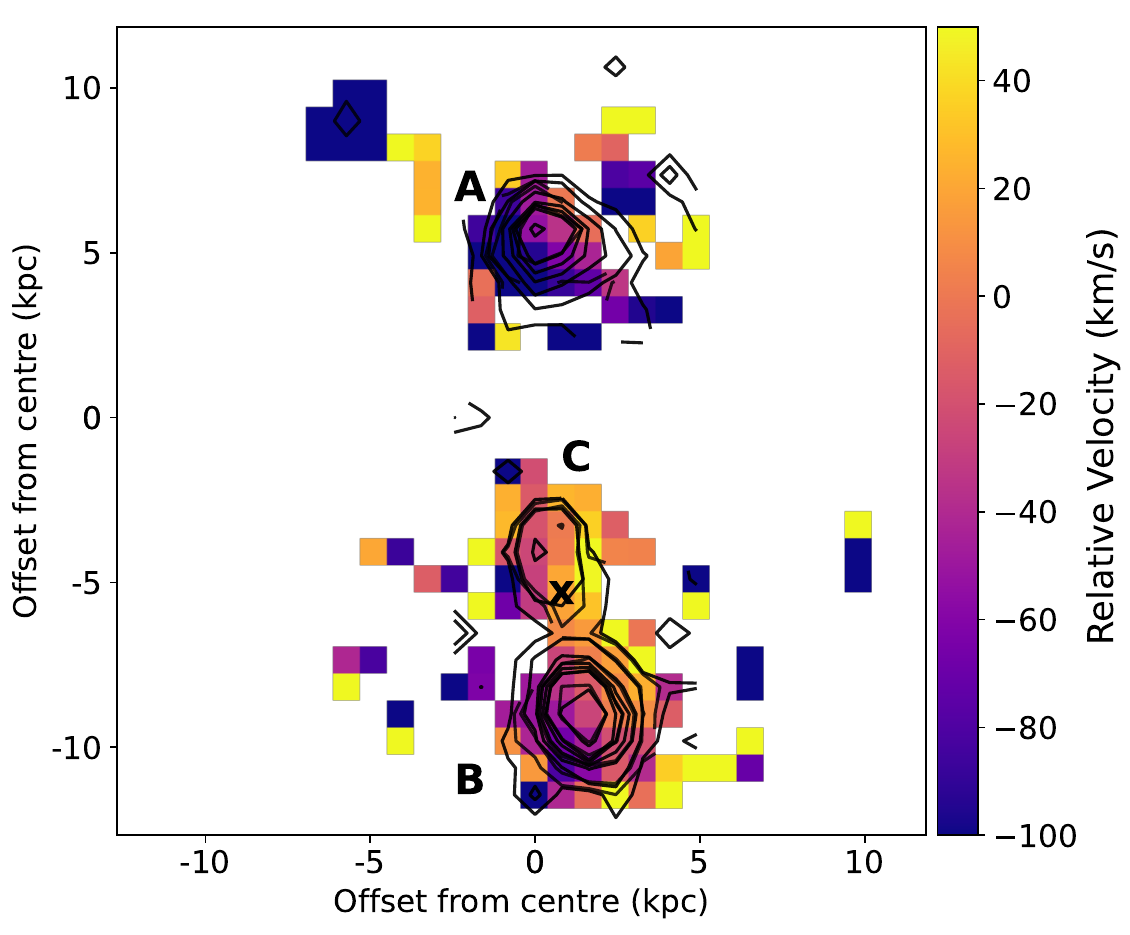}
    \end{subfigure}
    \begin{subfigure}
        \centering
        \includegraphics[width = \columnwidth]{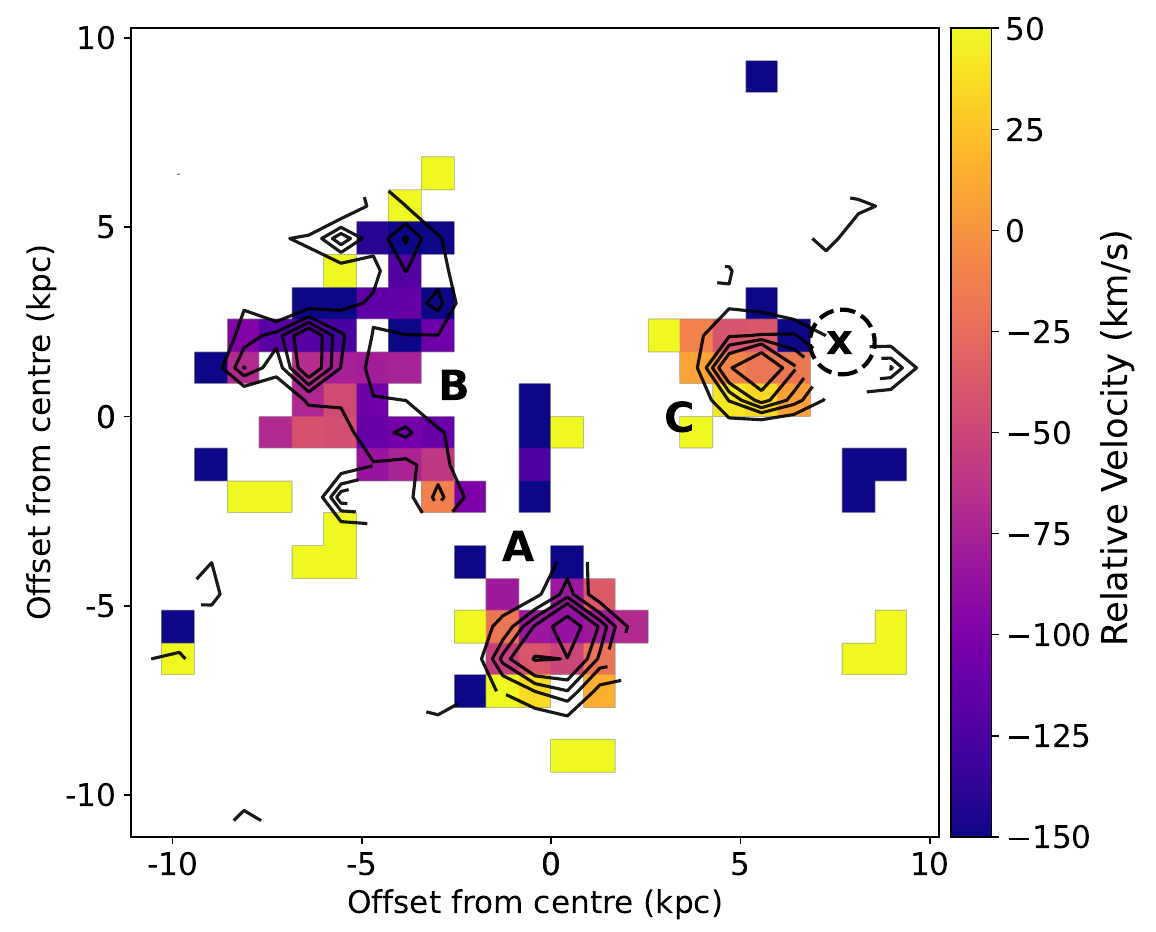}
    \end{subfigure}
    \caption{Maps showing the velocity offsets across the three detected star-forming components in GRB 050820A on the left panel and in GRB 150403A on the right relative to the GRB hosting component (i.e., component C in both). The contours trace the flux strength of [\ion{O}{iii}]$\lambda5007$ line emission as in Figures~\ref{fig:grb050820_flux} and \ref{fig:grb150403_o3flux}. Bright spaxels represent regions that are redshifted relative to the GRB hosting component and dark spaxels show blueshifted regions.}
    \label{fig:posmaps}
\end{figure*}

In order to study the relation between the multiple components detected in our IFS data we looked at the kinematics of each galaxy system. We created velocity field maps using the best-fit peak wavelength to the [\ion{O}{iii}]$\lambda5007$ line and converted this into rest-frame velocity space. Zero velocity was chosen to correspond to the redshift of the star-forming component hosting the GRB (component C in both cases). In order to reduce the noise present in the maps, a mask was applied to remove all spaxels with SNR $<2$. The velocity maps for the two GRB host galaxies are shown in Fig~\ref{fig:posmaps} where the contours corresponding to the three components identified in Figs.~\ref{fig:grb050820_flux} and \ref{fig:grb150403_o3flux} are overplotted in black.

In the data cube of GRB~050820A ( Fig.~\ref{fig:posmaps}, left panel), component A stands out with having the largest velocity difference, with a peculiar velocity of $-58 \pm 8$ km~s$^{-1}$ relative to component C. Although this is not a significant velocity offset, when combined with the projected separation of $\sim 8$~kpc from the other two components, the indication is that this is a separate but interacting galaxy in this system. In contrast, the spatial proximity of $< 3$~kpc and smaller velocity offset of $-26 \pm 8$ km~s$^{-1}$ between components B and C imply that these two components are interacting and possibly in the process of merging.

In the velocity map of GRB~150403A (Fig.~\ref{fig:posmaps}, right panel), both components A and B have velocity offsets relative to component C, indicating that they are blue-shifted relative to component C. Based on redshifts determined from stacked spectra, we measure peculiar velocities relative to component C of $-68 \pm 15$ km~s$^{-1}$ for component A and $-96 \pm 15$ km~s$^{-1}$ for component B. The projected separation between component C and the other two components ($\sim 10$~kpc) along with the velocity offsets, suggests that component C, which we associate with the GRB, is a separate galaxy. Despite having a separation of $\sim 20$~km/s and sharing similar rotational velocity axes, components A and B also appear to be separate galaxies. If components A and B were part of a single galaxy, the observed velocity field in Fig.~\ref{fig:posmaps} (right panel) would indicate rotation along the major axis, whereas galaxies are expected to rotate along the minor axis. Nonetheless, the close proximity of all three components in velocity space implies that all three are gravitationally interacting.

\begin{table*}
    \centering
    \caption{Total integrated and individual component emission line fluxes for all detected lines in the host galaxies of GRB 050820A and GRB150403A. All line fluxes have been corrected for Milky Way dust extinction and are presented here without host galaxy reddening correction. The [\ion{N}{ii}]$\lambda6584$ fluxes are reported as $1\sigma$ upper limits.}
    \label{tab:Host_fluxes}
    \begin{tabular}{lllllllll}
    \hline
    \multicolumn{1}{c}{\multirow{2}{*}{GRB Host}} & \multicolumn{1}{c}{\multirow{2}{*}{$z_{\mathrm{em}}^{*}$}} & \multicolumn{6}{c}{Line Flux ($10^{-17}$ erg cm$^{-2}$ s$^{-1}$)} \\ \cline{3-8}
     \multicolumn{1}{c}{} & \multicolumn{1}{c}{} & \multicolumn{1}{c}{H $\beta$} & \multicolumn{1}{c}{H $\alpha$} & \multicolumn{1}{c}{{[}\ion{O}{ii}{]}$\lambda\lambda3726,3729$} & \multicolumn{1}{c}{{[}\ion{O}{iii}{]}$\lambda4959$} & \multicolumn{1}{c}{{[}\ion{O}{iii}{]}$\lambda5007$} & \multicolumn{1}{c}{{[}\ion{N}{ii}{]}$\lambda6584$}\\ \hline 
     050820A & & & & & & \\ \hspace{3mm}total integrated & $2.6131$ & $2.85 \pm 0.11$ & $10.49 \pm 0.26$ & $5.13 \pm 0.22$ & $6.33 \pm 0.17$ & $18.66 \pm 0.19$ & $< 0.80$\\
    \hspace{3mm}component A & $2.6129$ & $1.14 \pm 0.08$  & $4.67 \pm 0.13$ & $1.76 \pm 0.12$ & $2.43 \pm 0.11$ & $7.17 \pm 0.15$ & $< 0.41$ \\
    \hspace{3mm}component B & $2.6133$ & $1.50 \pm 0.05$ & $4.83 \pm 0.27$ & $2.38 \pm 0.07$ & $3.35 \pm 0.08$ & $9.87 \pm 0.11$ & $<0.39$ \\
    \hspace{3mm}component C & $2.6136$ & $0.42 \pm 0.04$ & $1.44 \pm 0.21$ & $0.98 \pm 0.06$ & $0.68 \pm 0.10$ & $2.00 \pm 0.12$ & $<0.12$  \\
    150403A & & & & & &\\ \hspace{3mm}total integrated & $2.0563$ & $2.88 \pm 1.43$ & $10.64 \pm 3.18$ & $7.57 \pm 2.71$ & $4.34 \pm 1.68$ & $12.78 \pm 2.36$ & $<2.96$ \\ \hspace{3mm}component A & $2.0564$ & $1.62 \pm 0.29$ & $2.80 \pm 0.52$ & $1.74 \pm 0.50$ & $1.28 \pm 0.31$ & $3.78 \pm 0.44$ & $<0.19$\\
    \hspace{3mm}component B & $2.0562$ & $2.17 \pm 0.52$ & $5.76 \pm 1.26$ & $4.62 \pm 0.90$ & $2.32 \pm 0.55$ & $6.82 \pm 0.77$ & $<0.69$\\
    \hspace{3mm}component C & $2.0571$ & $0.14 \pm 0.21$ & $1.71 \pm 0.68$ & $0.94 \pm 0.35$ & $1.16 \pm 0.29$ & $3.40 \pm 0.40$ & $<0.16$ \\ \hline
    \multicolumn{8}{l}{\footnotesize $^*$ In all cases the measured redshift uncertainty is 0.0002.}
    \end{tabular}
\end{table*}

\subsection{Star formation rates}
The SFR of each component is estimated using the dust corrected integrated $\mathrm{H}\alpha$ fluxes \citep{kennicutt1998_HaSFR} assuming a \cite{chabrier2003_IMF} initial mass function. The estimated values are reported in Table \ref{tab:Metal_SFR}. For GRB 050820A, component C, which we associate with the GRB, has the lowest SFR. It is nevertheless still clearly a star-forming region, with an H$\alpha$ SFR of $5.4 \pm 2.0 ~\mathrm{M}_{\odot} \mathrm{yr}^{-1}$. GRB 150403A also lies close to a star-forming region (component C), although SFR is more uncertain due to the undetected H$\beta$ line. The H$\alpha$ SFR of component C is $39^{+76}_{-36} ~\mathrm{M}_{\odot} \mathrm{yr}^{-1}$. Here, the lower limit corresponds to the SFR inferred from the dust uncorrected H$\alpha$ flux, corresponding to $\sim 3 ~\mathrm{M}_{\odot}\mathrm{yr}^{-1}$. From Fig.~\ref{fig:grb150403_o3flux} we measure a projected separation between component C and the GRB position that is less than $\sim 1.6 \mathrm{kpc}$ when accounting for the GRB positional uncertainty. This is consistent within 1$\sigma$ of the median offset measured in \cite{blanchard2016_GRBoffset} and therefore suggests that while the GRB did not occur in the region of peak star formation, it is still likely associated with the nearby star-forming component.

\subsection{Emission line metallicities}
\begin{figure*}
    \centering
    \includegraphics[width = 0.99\textwidth]{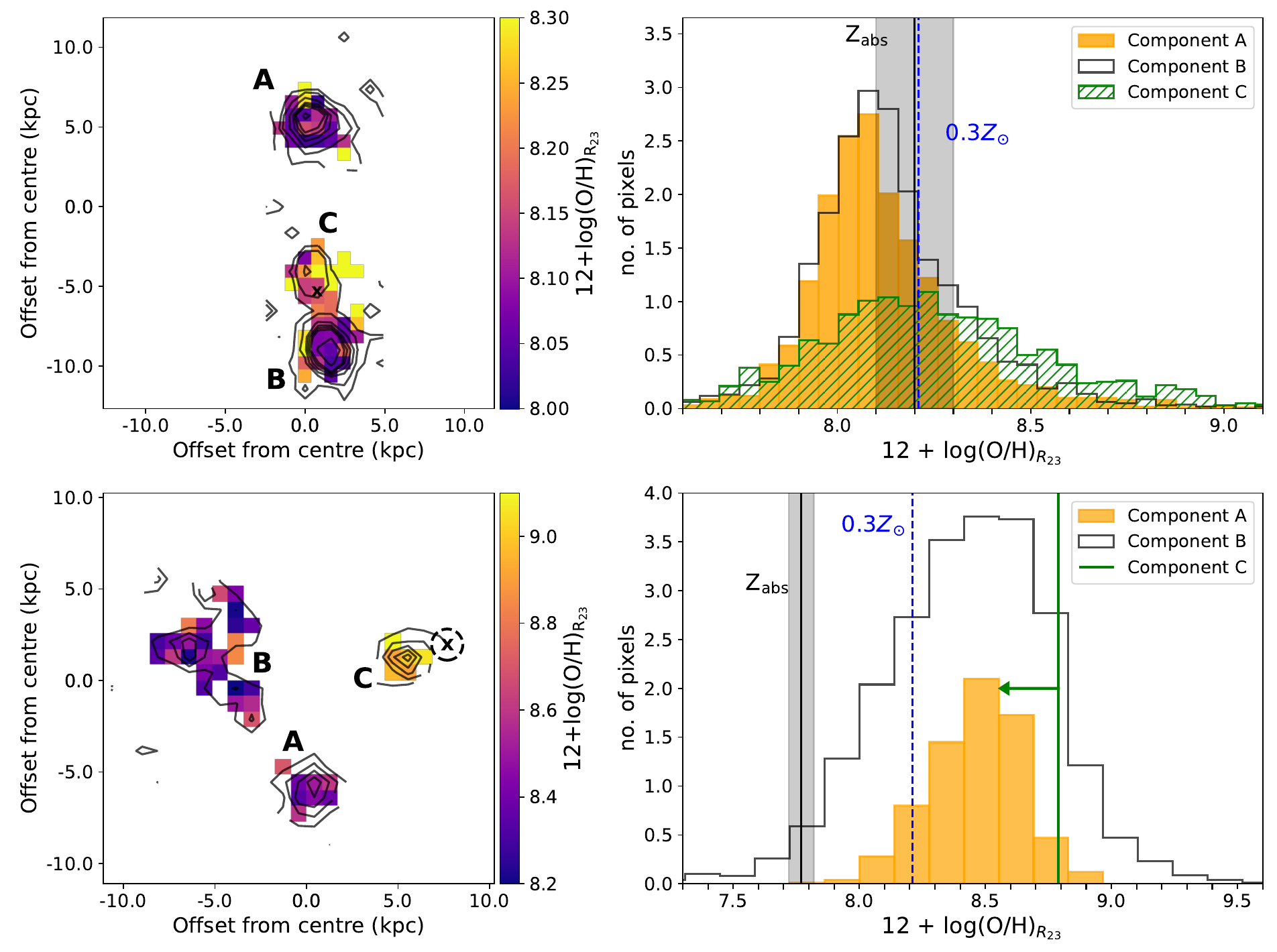}
    \caption{Metallicity variation across the two GRB host galaxies. The top row corresponds to GRB 050820A, and the bottom row to GRB 150403A. \textit{Left}: Gas-phase metallicity derived using the R$_{23}$ diagnostic from \protect\cite{nakajima2022NOX22}, with contours tracing [\ion{O}{iii}]$\lambda5007$ flux. For GRB 150403A component C, the metallicities are represented using $1\sigma$ upper limits. \textit{Right}: Metallicity distributions of components A (filled orange), B (black outline), and C (hatched green), from the metallicity maps obtained by Monte Carlo sampling from the metallicity maps. The solid, black vertical lines indicate the GRB afterglow absorption metallicity, Z$_{\rm abs}$, converted to an equivalent oxygen abundance with its $1\sigma$ uncertainty represented by the shaded region. Finally, the theoretical metallicity cap of $0.3Z_{\odot}$ predicted by the single star collapsar model is indicated by the dashed, vertical blue line. In the bottom right panel, the upper limit metallicity on component C is represented by the vertical green line and the leftwards arrow.}
    \label{fig:R23_panel}
\end{figure*}

\begin{figure*}
    \centering
    \includegraphics[width = 0.99\textwidth]{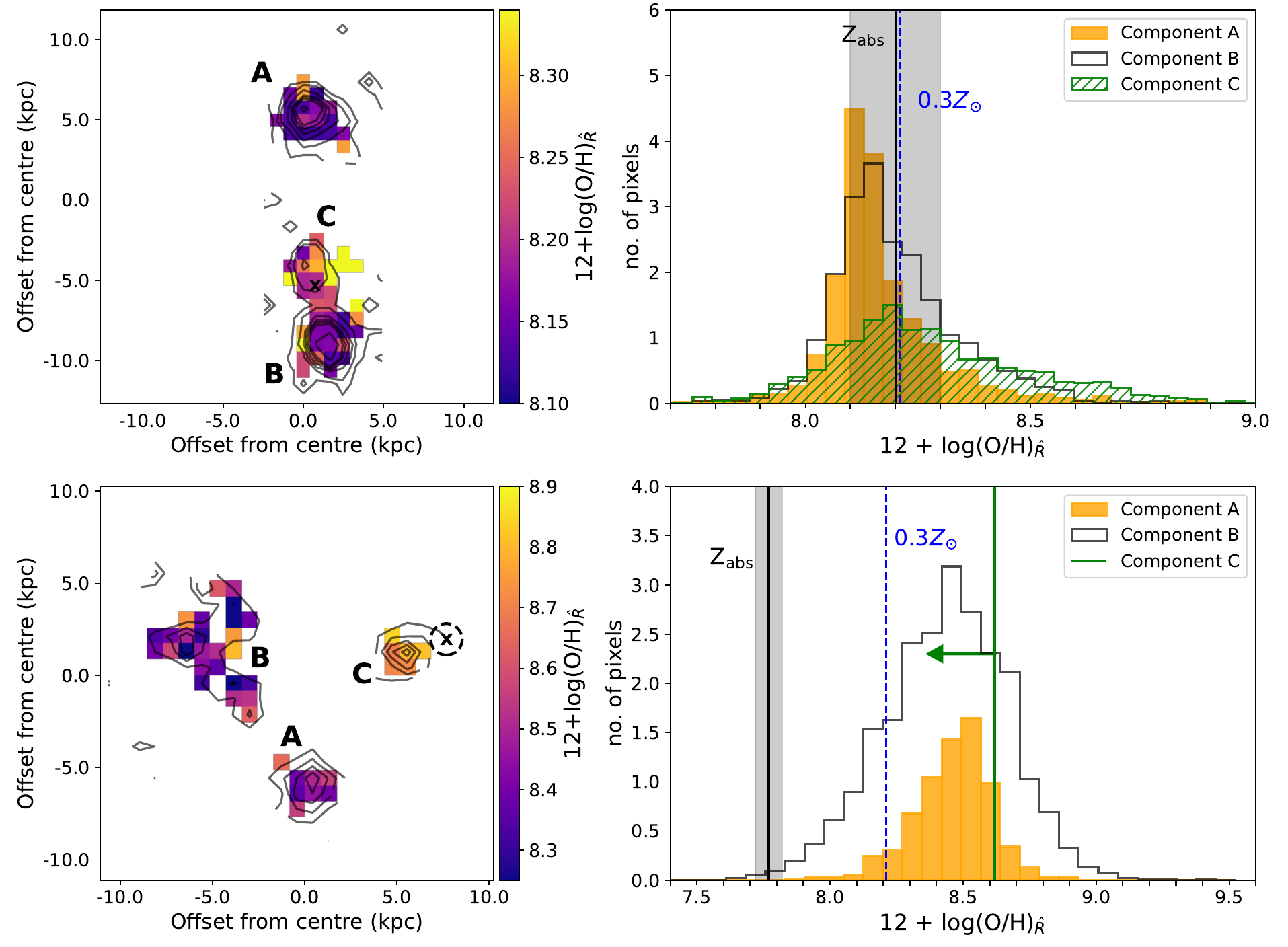}
    \caption{Figure showing the $\hat{R}$ metallicity maps from \protect\cite{laseter2024_Rhat} on the left, and the histograms representing the distribution of metallicities across the three components on the right for GRB 050820A (top row) and GRB 150403A (bottom row).}
    \label{fig:Rhat_panel}
\end{figure*}

Multiple gas-phase metallicity line diagnostics, based on different emission line ratios, are available, each with their own advantages and disadvantages. We chose to consider diagnostics that are calibrated using high-$z$ galaxies or their low-redshift analogues, and that are least sensitive to the effects of ionisation. These are the $\hat{R}$ diagnostic from \cite{laseter2024_Rhat}, and the $\mathrm{R_{23}}$ diagnostics from \cite{nakajima2022NOX22} and \cite{sanders2024SST23}. In addition to calibrating multiple diagnostics on their full galaxy sample, \cite{nakajima2022NOX22} also provide H$\beta$ equivalent width (EW)-dependent calibrations for their low metallicity sample ($12+\log({\rm O/H})\sim 8.0$). However, we unfortunately do not detect the galaxy continuum in our NIRSpec data, and are therefore unable to measure H$\beta$ EW. In this work, we therefore adopt the diagnostic calibrated using the full available sample.
From the dust corrected line fluxes,  we calculate the diagnostic line ratios R$_{2}$=log([\ion{O}{ii}]$\lambda\lambda3726,3729$/H$\beta$), R$_{3}$=log([\ion{O}{iii}]$\lambda5007$/H$\beta$), R$_{23}$=log(([\ion{O}{ii}]$\lambda\lambda3726,3729$+[\ion{O}{iii}]$\lambda\lambda4959,5007$)/H$\beta$), and $\hat{R}$ = 0.47 R$_{2}$ + 0.88 R$_{3}$ from \cite{laseter2024_Rhat}.

\subsubsection{Selecting the appropriate metallicity branch}\label{sec:branch}
A drawback of all three diagnostics used here is that they are double branched with a turnover that can yield two metallicity solutions for a given line ratio. However, for the $\mathrm{R}_{23}$ values we measure, the \cite{sanders2024SST23} $\mathrm{R}_{23}$ diagnostic has no valid upper-branch solutions due to its shorter validity range of metallicities. Therefore, we exclude \cite{sanders2024SST23} $\mathrm{R}_{23}$ from the reported metallicity results. Nevertheless, we provide the lower-branch solutions in Table~\ref{tab:SST24_lower}.

The turnover for the \cite{laseter2024_Rhat} $\hat{R}$ diagnostic is at 12+log(O/H) = 8.12, while for the two $\mathrm{R_{23}}$ diagnostics, it is approximately 8.00. In order to determine whether to use the lower or upper branch solutions, we considered several additional factors for both galaxies.
First, solutions calculated in this work were compared to the GRB afterglow absorption line metallicities. For GRB 050820A, \cite{wiseman2017abs_metallicity} reported a sightline host galaxy metallicity $\mbox{[M/H]}=-0.49\pm 0.10$\footnote{$\mbox{[M/H]}=\frac{\log N_M}{\log N_H}-\frac{\log N_{M_\odot}}{\log N_{H_\odot}}$ for some element $M$}, which corresponds to an oxygen abundance $12+\log({\rm O/H})=8.20 \pm 0.10$. For GRB 150403A, the absorption metallicity was $\mbox{[M/H]}=-0.92\pm 0.05$ \citep{bolmer2019_150403abs_Z}, corresponding to $12+\log({\rm O/H})=7.77 \pm 0.05$. The absorption line metallicities along the GRB line of sight thus suggest an upper-branch solution for GRB 050820A and a lower-branch for GRB 150403A. However, it remains unclear whether we should expect the absorption line metallicity, which probes the galaxy neutral gas, to be the same as the ionised gas within star-forming regions that we probe here with nebular emission lines \citep{schady2024comparing}. 

We therefore also considered additional strong line diagnostics to help break the degeneracies between the two-branch solutions. Following the same procedure as in \cite{curti+20}, we combined the R$_{23}$, R$_{3}$ and O32 diagnostics from \cite{nakajima2022NOX22} into a single estimate and determined the metallicity that minimised the difference between the observed and predicted diagnostic line ratio, taking into account the uncertainties in the line ratios. Using this technique, we measure a metallicity of $8.28 \pm 0.12$ for GRB 150403A component C and $8.10 \pm 0.10$ for GRB 050820A component C, suggesting that for both host galaxies the upper-branch solution provides the greatest mutual consistency with all line diagnostics considered here. It is worth noting that the result from this analysis in the case of GRB~150403A is strongly influenced by the O32 diagnostic, as it has the smallest uncertainty due to not relying on the H$\beta$ line. The O32 diagnostic is known to be highly sensitive to ionisation parameters \citep[e.g.][]{kewley2002_stronglineeg}, which adds additional systematic uncertainty to the metallicity measured in component C of GRB~150403A.

Finally, we also consider the galaxy metallicity expected from scaling relations given by other measured characteristic properties. In the case of GRB 050820A, we measure a star formation rate (SFR) of $5.4 \pm 2.0 ~\mathrm{M}_{\odot} \mathrm{yr}^{-1}$ in component C, and use the stellar mass of $\log(M_{\star} / M_{\odot})$ = 9.29 for the same component measured in \cite{chen2009_GRB050820}. Using the mass metallicity relation (MZR) from \cite{sanders2021FMR_MZR}, we would expect a component C metallicity of $12+\log({\rm O/H})=8.30\pm 0.03$. Similarly, when considering the SFR-dependence on the MZR described by the fundamental metallicity relation (FMR), we would expect a comparable metallicity of $12+\log({\rm O/H})=8.37 \pm 0.05$ based on the high-z FMR from \cite{sanders2021FMR_MZR}. The results obtained again support an upper-branch solution for component C. Similar outcomes were obtained for the other components after performing the same analysis and therefore we report the upper-branch solutions in Table~\ref{tab:Metal_SFR}. It is worth noting one caveat with this method, which is that \cite{sanders2021FMR_MZR} use the `B18' \citep{bian2018_B18_diagnostic} metallicity diagnostic in their determinations of the MZR and FMR that is also based on strong-line calibrations.

For GRB 150403A component C, we only have a lower limit on the SFR due to our uncertain dust correction. We therefore refrain from making a similar analysis as in the case of GRB 050820A. We are, however, able to use the measured SFRs of components A and B, and, assuming they are main sequence galaxies, combine these with the galaxy main sequence (SFR-$M_{\star}$ relation) from \cite{sanders2021FMR_MZR} to obtain a stellar mass estimate of $\log(M_{\star} / M_{\odot}) = 8.91 \pm 0.21$ and $9.33 \pm 0.17$, respectively. For these masses the \citet{sanders2021FMR_MZR} MZR predicts a metallicity 12+log(O/H) = $8.22 \pm 0.11$ for component A and $8.33 \pm 0.09$ for component B. However, it is also worth noting that there is a large systematic uncertainty associated to these metallicity estimates given the large scatter in both the galaxy main sequence and the MZR.

Although the relations used within the methods above rely on various assumptions and empirical relations, all estimates consistently indicate an upper-branch solution with the exception of the absorption metallicity of GRB 150403A. We therefore adopt the upper branch solution for both host galaxies. 

\subsubsection{Metallicity distribution}
In addition to reporting the total integrated and component-based $\hat{R}$ and R$_{23}$ metallicities, and line ratios in Table~\ref{tab:Metal_SFR}, as was done in \cite{schady2024comparing}, we also determine the metallicity at each IFU pixel to produce metallicity maps and study the metallicity variations across the two systems. In Fig.~\ref{fig:R23_panel}, we show the \cite{nakajima2022NOX22} R$_{23}$ metallicity maps in the left panel for the host galaxies of GRB 050820A (top row) and GRB 150403A (bottom row). Similarly, in Fig.~\ref{fig:Rhat_panel}, we show the \cite{laseter2024_Rhat} $\hat{R}$ metallicity maps for the two host galaxies. For the metallicity maps, the same masking criteria as in Fig.~\ref{fig:posmaps} is applied, removing all spaxels with [\ion{O}{iii}]$\lambda5007$ SNR $< 2$. Additionally, in Figs.~\ref{fig:R23_panel} and \ref{fig:Rhat_panel}, we also mask those spaxels for which our H$\beta$ fits failed. In addition to the metallicity maps, we also show the metallicity distribution within each of the components in the right panels of both figures. To take into account the measurement metallicity uncertainties on the generated histogram, we applied a Monte Carlo (MC) sampling, where we randomly selected the metallicity of each pixel from a Gaussian distribution centred at the measured pixel metallicity with a width corresponding to the metallicity $1\sigma$ uncertainty. We ran this 100 times, producing a histogram after each run, and the average of these 100 realizations is then used to generate an MC-averaged histogram for each component. The filled orange, solid black and hashed green histograms in the right panels of Fig.~\ref{fig:R23_panel} and Fig.~\ref{fig:Rhat_panel} correspond to the results from our MC analysis for components A, B and C, respectively. Note that for GRB 150403A (bottom panels), a green vertical line with a leftwards arrow is used to represent the upper limit placed on the measured metallicity value instead of a histogram. On the same histograms, the GRB afterglow absorption metallicity is indicated with the labelled solid, black vertical line along with its $1\sigma$ uncertainty in shaded regions. Finally, the $0.3Z_\odot$ metallicity cap predicted by the long GRB collapsar model is indicated with a blue, vertical dashed line.

The metallicity maps for GRB 050820A (Fig.~\ref{fig:R23_panel} and \ref{fig:Rhat_panel}, top left panels) show no obvious patterns of variation or gradients. However, components A and B exhibit a slightly inverted metallicity gradient, with lower metallicities in the centres and higher metallicities in the outskirts. In contrast, the histograms in the top panels of Figs.~\ref{fig:R23_panel} and \ref{fig:Rhat_panel} indicate that the metallicity distributions across the three components are largely consistent, apart from a few outlier spaxels. 

For the likely host galaxy of GRB 050820A (component C), we measure metallicities of $8.24 \pm 0.25$ and $8.27 \pm 0.18$ using the \cite{nakajima2022NOX22} R$_{23}$ and \cite{laseter2024_Rhat} $\hat{R}$ diagnostics, respectively. From the IFU data, we also determine the metallicity at the GRB location to be $8.17 \pm 0.22$ and $8.19 \pm 0.15$ for the \cite{nakajima2022NOX22} R$_{23}$ and \cite{laseter2024_Rhat} $\hat{R}$ diagnostics, respectively. All four measurements are consistent with the absorption metallicity, which is equivalent to an oxygen abundance of $12+\log({\rm O/H})=8.20 \pm 0.10$ \citep{wiseman2017abs_metallicity}. This can also be seen in the right-most top panels of Figs.~\ref{fig:R23_panel} and \ref{fig:Rhat_panel}, where the vertical line indicating the absorption metallicity, Z$_{\rm abs}$, lies within the metallicity distribution of all three components. When compared to the maximum metallicity of long GRB progenitor stars predicted by the single star collapsar model, the histograms of all three components peak at metallicities below 0.3Z$_\odot$, although it is noteworthy that component C, which we associate with the GRB, shows a broader distribution than components A and B. Nevertheless, the measured metallicity is consistent with being below $0.3Z_{\odot}$ when considering uncertainties. 

For GRB 150403A, the metallicity maps show no clear patterns. From the histograms on the bottom panels, all three components appear to be inconsistent ($\sim 0.6 ~\mathrm{dex}$ higher) with the measured absorption metallicity of $7.77 \pm 0.05$ for GRB 150403A \citep{bolmer2019_150403abs_Z}. For component C we only indicate the $1\sigma$ upper limit on the measured metallicity due to the non-detection of the H$\beta$ line for this component. The corresponding upper limits we measure are $<8.79$ using the \cite{nakajima2022NOX22} R$_{23}$ diagnostic and $< 8.62$ using the \cite{laseter2024_Rhat} $\hat{R}$ diagnostic.

We also note that, a new metallicity calibration was published by \cite{Scholte_SCC25}, providing a recalibration of the \cite{laseter2024_Rhat} $\hat{R}$ diagnostic using a larger sample of high-redshift galaxies. We tested our results with the updated $\hat{R}$ calibration and found them to be consistent with the \cite{laseter2024_Rhat} $\hat{R}$ values reported in Table \ref{tab:Metal_SFR}, within $1\sigma$.

\subsection{The ISM Properties}
\begin{figure*}
    \centering
    \includegraphics[width = \textwidth]{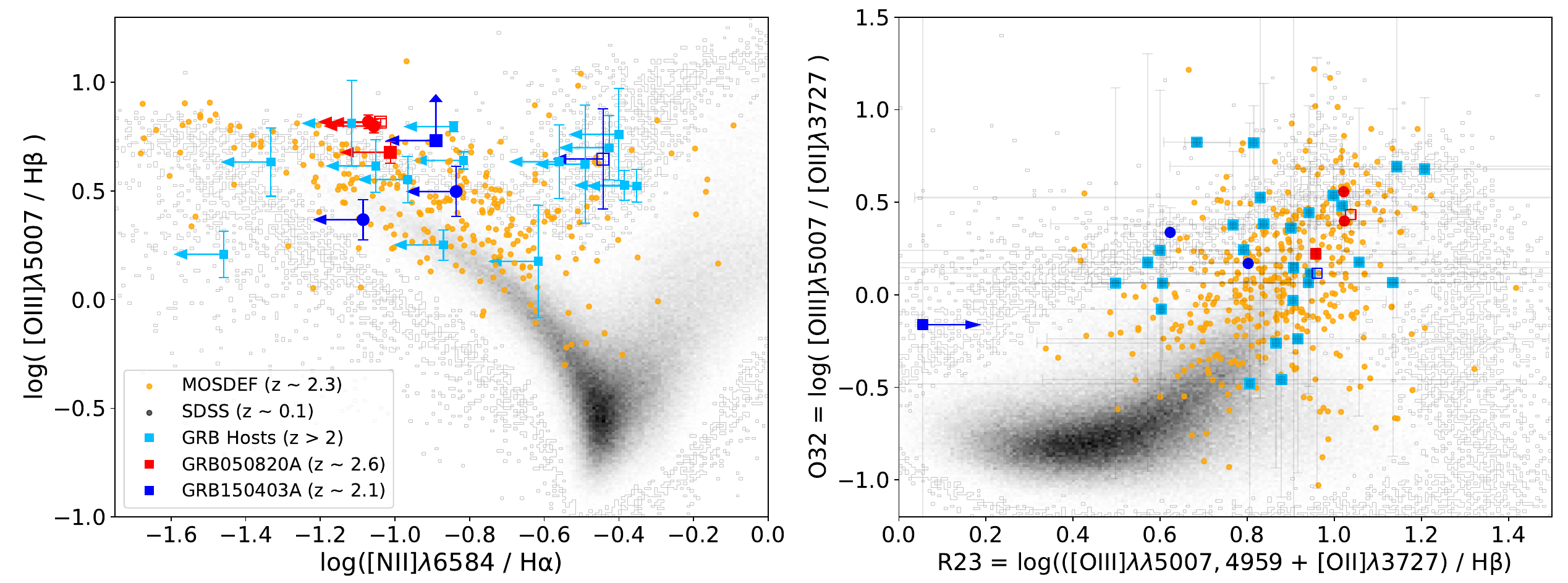}
    \caption{\textit{Left}: The N2-BPT diagram with $1\sigma$ upper limits on [\ion{N}{ii}]/H$\alpha$. \textit{Right}: The O32 vs R23 diagram. In both diagrams the star-forming components presented in this work are plotted in red (for GRB 050820A) and blue (GRB 150403A) with the likely host galaxies (component C) represented with filled squares and the galaxy-integrated measurements with empty squares. Finally, the remaining two star-forming components from both observations (components A and B) are represented as blue and red circles. Note that for GRB 150403A component C, additional $1\sigma$ lower limits are placed on line ratios containing H$\beta$. The cyan data points represent a sample of high redshift ($z > 2$) GRB hosts observed with {X-shooter} \citep{kruhler2015grb}, and with {\em JWST}/NIRSpec in fixed slit mode \citep{schady2024comparing}, including the host galaxy of GRB 050505 at $z=4.27$ (Inkenhaag et al., in prep.). Orange data points are high-redshift ($z \sim 2.3$) star-forming galaxies from the MOSDEF survey while the black data points represent local star-forming galaxies from the SDSS catalogue.}
    \label{fig:BPT}
\end{figure*}
The Baldwin, Phillips \& Terlevich (BPT) diagram of $\log([\ion{O}{iii}]\lambda 5007/H\beta)$ against $\log([\ion{N}{ii}]\lambda 6584/H\alpha)$ is effective at identifying the dominant ionisation mechanism responsible for the observed emission line, and it highlights differences in the ionisation properties of local and distant galaxies  \citep{bptdiagram}. The O32 (=log([\ion{O}{iii}]$\lambda5007$ / [\ion{O}{ii}]$\lambda\lambda3726,3729$)) vs R$_{23}$ diagram similarly depicts the ISM conditions of the nebular gas \citep[e.g.,][]{nakajima2014_O32_R23, paalvast2018_R23_O32, runco2021mosdef_O32vsR23}, with R23 tracing metallicity and O32 tracing ionisation. The BPT diagram and the O32 vs R23 diagrams in Fig.~\ref{fig:BPT} show the emission line properties for each of the three components identified in our two likely host galaxies as filed squares as well as the total integrated estimates as empty squares. The remaining two star-forming components are represented with circles. For comparison, we also show a sample of 15 high redshift ($z > 2$) GRB host galaxies on both diagrams as the cyan data points from \cite{kruhler2015grb} and \cite{schady2024comparing} as well as GRB 050505 observed with {\em JWST} at $z \sim 4.3$ (Inkenhaag et al., in prep.). For the majority of GRB host galaxies [\ion{N}{ii}]$\lambda6584$ is undetected, and we thus use the $1\sigma$ upper limit, which is indicated by the leftward arrows. For GRB 150403A component C, additional $1\sigma$ lower limits are shown for the $\log([\ion{O}{iii}]\lambda 5007/H\beta)$ and R23 line ratios due to the non-detection of H$\beta$. Gray data points represent the local star-forming galaxies from the seventh data release of the SDSS survey \citep{abazajian2009SDSS_DR7} while orange data points represent high-redshift ($z \sim 2.3$) star-forming galaxies from MOSDEF \citep{kriek2014_MOSDEF1, reddy2015_MOSDEF2}. Only SDSS and MOSDEF galaxies with H$\beta$, [\ion{O}{ii}]$\lambda\lambda3726,3729$ and [\ion{O}{iii}]$\lambda5007$ fluxes with $\mathrm{SNR} > 3$ are plotted.

High redshift star-forming galaxies are expected to shift upwards on the BPT diagram compared to local galaxies due to the harder ionisation fields and higher ionisation parameters at higher redshift \citep{kewley2013, shirazi2014_BPT_ionization, steidel2014_BPTdashed}. This behaviour is observed in the MOSDEF galaxy at $\mathrm{z} \sim 2.3$ (orange) and the $z>2$ GRB host galaxy sample from \cite{kruhler2015grb} and \cite{schady2024comparing} (cyan), which lie systematically above the SDSS galaxies (Fig.~\ref{fig:BPT} left panel).  For GRB 050820A, all three components, as well as the total integrated data point, appear relatively consistent with one another. For GRB 150403A, components A and B appear consistent with the distribution of high redshift GRB hosts and galaxies. In the case of component C, we only have limits on the line ratios due to the non-detection of [\ion{N}{ii}]$\lambda6584$ and H$\beta$. Nevertheless, the star-forming components studied in this work lie in a region of the BPT diagram consistent with other GRB host and star-forming galaxies at $z>2$. In the right panel of Fig.~\ref{fig:BPT} we show the O32 vs R23 line ratios for the same sample of galaxies as plotted in the BPT diagram. We again see an offset in the $z>2$ galaxies relative to the SDSS sample, with high-redshift galaxies clustering around the tail end of the distribution of local SDSS galaxies, corresponding to the generally higher ionisation fields and lower abundances in high redshift galaxies as already seen in the BPT diagram \citep[e.g,][]{shapley2015mosdef}. In this parameter space our sample of star-forming components again lie within the same region of the parameter space as occupied by other high-redshift GRB hosts and star-forming galaxies. This indicates that these GRB host galaxies are not particularly unusual or distinct compared to high-z star-forming galaxies.

\begin{table*}
    \caption{Table with NOX22 R$_{23}$ \citep{nakajima2022NOX22} and LMC24 $\hat{R}$ \citep{laseter2024_Rhat} emission line metallicities and H$\alpha$ SFRs of the total integrated and individual components detected in the host galaxies of GRB 050820A and GRB 150403A. The measured host galaxy $E(B-V)$ and $\sigma$ values are also shown.}
    \label{tab:Metal_SFR}
    \begin{center}
    \begin{tabular}{llllllll}
    \hline
    \multicolumn{1}{c}{\multirow{2}{*}{GRB Host}} & \multicolumn{1}{c}{\multirow{2}{*}{$\hat{R}$}}& \multicolumn{1}{c}{\multirow{2}{*}{R$_{23}$}} & \multicolumn{2}{c}{12+log(O/H)} & \multicolumn{1}{c}{SFR$_{\mathrm{H}\alpha}$} & \multicolumn{1}{c}{$E(B-V)$} & \multicolumn{1}{c}{$\sigma$}\\ \cline{4-5} 
    \multicolumn{1}{c}{} & \multicolumn{1}{c}{} & \multicolumn{1}{c}{} & \multicolumn{1}{c}{NOX22 R$_{23}$} & \multicolumn{1}{c}{LMC24 $\hat{R}$} & \multicolumn{1}{c}{($M_{\odot}$/yr)} & \multicolumn{1}{c}{(mag)} & \multicolumn{1}{c}{(km~s$^{-1}$)} \\ \hline 
     050820A & & & & & & &\\ 
     \hspace{3mm}total integrated & $0.88 \pm 0.10 $ & $1.04 \pm 0.07 $& $8.04 \pm 0.07$ & $8.13 \pm 0.04$ & $46 \pm 4$ & $0.23 \pm 0.04$ & $29 \pm 5$\\
    \hspace{3mm}component A & $0.87 \pm 0.17$ & $1.02 \pm 0.12$ & $8.10 \pm 0.14$ & $8.17 \pm 0.09$ & $25 \pm 4$ & $0.33 \pm 0.07$ & $26 \pm 9$ \\
    \hspace{3mm}component B & $0.84 \pm 0.12$ & $1.02 \pm 0.10$ &$8.09 \pm 0.12$ & $8.17 \pm 0.09$ & $16 \pm 2$ & $0.11 \pm 0.06$ & $35 \pm 4$\\
    \hspace{3mm}component C & $0.80 \pm 0.38$ & $0.96 \pm 0.27$ &$8.24 \pm 0.25$ & $8.27 \pm 0.18$ & $5.4 \pm 2.0$ & $0.16 \pm 0.16$ & $< 98$\\
    150403A & & & & & & & \\ 
    \hspace{3mm}total integrated & $0.81 \pm 1.22$ & $0.96 \pm 0.87$ &$8.40 \pm 0.42$ & $8.39 \pm 0.31$ & $24^{+27}_{-8}$ & $0.20^{+0.50}_{-0.20}$ & $< 116$\\ 
    \hspace{3mm}component A & $0.34 \pm 0.13$ & $0.62 \pm 0.09$ & $8.60 \pm 0.08$ & $8.57 \pm 0.05$ & $4.2\pm 0.8$ & $< 0.25^*$ & $56 \pm 39$ \\
    \hspace{3mm}component B & $0.59 \pm 0.15$ & $0.80 \pm 0.11$ & $8.41 \pm 0.16$ & $8.40 \pm 0.12$ & $8.7 \pm 1.9$ & $ < 0.83^* $ & $< 116$\\
    \hspace{3mm}component C & $1.86 \pm 2.31$ & $1.77 \pm 1.71$ &$< 8.79^{\dagger}$ & $< 8.62^{\dagger}$ & $39^{+76}_{-36}$ & $1.35 \pm 0.94$ & $< 116$\\ \hline 
    \end{tabular}
    \end{center}
    
\footnotesize{$^*$ In these cases the measured E(B-V) was less than zero, so we report the corresponding $3\sigma$ upper limits and use E(B-V) = 0 in our calculations. \\
$^\dagger$ The metallicity of GRB 150403A component C is reported with $1\sigma$ upper limits due to the non-detection of H$\beta$ for this component.}
\end{table*}

\section{Discussion}\label{sec:discussion}
\subsection{Relation between components}
In the host galaxy observations of both GRB~050820A and GRB~150403A three spatially distinct emission components are clearly visible. Aside from their projected separation, the IFU data provide key emission line properties (section~\ref{sec:result}) which we use to determine the relation between the detected components and to the GRB itself. We have labelled these components A, B and C, and in both cases the GRB position is spatially located closest to component C. 

\subsubsection{GRB~050820A}
Within the galaxy complex of GRB 050820A, component A is well separated from the other two components, with a projected physical separation of $\sim 8$kpc \cite[Fig~\ref{fig:grb050820_flux}; see also][]{chen2009_GRB050820}. This separation is $\sim$ 4 times larger than the average size of star-forming galaxies at similar redshifts \citep{ribeiro2016_size_vs_z}, suggesting that component A is a separate galaxy to the GRB host, as already concluded in \cite{chen2009_GRB050820}. On the other hand, components B and C exhibit a smaller velocity offset ($-26 \pm 8$ km~s$^{-1}$), and their separation of less than 3kpc is more consistent with them being part of a single galaxy. Additionally, emission is observed across components B and C in continuum as detected in the {\em HST} broadband imaging, as well as in our [\ion{O}{iii}]$\lambda5007$ surface brightness maps (see Fig.~\ref{fig:grb050820_flux}). However, interpreting them as a single system would imply a total diameter of $\sim 8$kpc, which, as stated above, would be far larger than the average size of star forming galaxies at a similar redshift. Furthermore, the morphology and dynamics of the system do not appear disk-like (see Fig.~\ref{fig:grb050820_flux}), making it more likely that components B and C are also separate galaxies. 

We can use the velocity offsets measured between the three components, combined with the GRB afterglow absorption spectra from \cite{prochaska2007_GRB050820absredsft} to build a 3D picture of the system. The absorption spectra reveal broad absorption features spanning up to $\Delta v \sim 400 ~\mathrm{km s^{-1}}$, but due to degeneracies between local velocities and cosmological redshift, we are unable to robustly identify the origin of the high velocity absorbing material. As described in \cite{chen2012_GRB050820A_HST}, either component C lies in front of components A and B and is falling towards them, or it is located behind A and B but moving away from component B. In our NIRSpec IFS data we additionally measure a velocity offset of $-97 \pm 17 ~\mathrm{km s^{-1}}$ for component C relative to the GRB absorption redshift of 2.61469 \citep{prochaska2007_GRB050820absredsft}. Given the projected proximity between the GRB and component C, it is natural to assume that component C is the host galaxy, and the afterglow spectrum shows blueshifted absorption features consistent with the velocity offset measured to component C. This would therefore imply that turbulence within the host galaxy (i.e. component C) contributes to the velocity spread of $\sim 100$~km/s observed in the afterglow spectrum, with additional absorption at larger velocities originating from material within B and in tidal debris between components A and B, as suggested in \cite{chen2012_GRB050820A_HST}.

An additional reason to favour a scenario in which component C is the GRB host galaxy and lies behind components A and B is the detection of the excited line \ion{Si}{ii}*$\lambda1264$ absorption feature in the GRB absorption spectrum consistent with the velocity of component B. \ion{Si}{ii}*$\lambda1264$ is an excited-state transition and previous time evolution observations of excited lines in GRB afterglow spectra have indicated that the GRB is the source of excitation, placing the absorbing material within a few kpc of the GRB \citep{vreeswijk2007_excited_state, d2010_excited_090926A, vreeswijk2013_excited_state, saccardi2023_excited_210905A}. This is consistent with the measured separation between components B and C, which may help break the degeneracy between the two possible scenarios, ultimately favouring the interpretation that the GRB host lies behind the other two interacting galaxies.

\subsubsection{GRB~150403A}
Our IFU observations of the host galaxy of GRB 150403A reveal a significant projected separation ($\sim 7$ kpc) between the component likely to have hosted the GRB (component C) and the other two components (see Fig.~\ref{fig:grb150403_o3flux}). This separation suggests that component C is a distinct galaxy that is interacting with components A and B, which are offset from C by $-68 \pm 15$ km~s$^{-1}$ and $-96 \pm 15$ km~s$^{-1}$, respectively.
The smaller velocity offset between A and B ($\sim 28~\mathrm{km~s^{-1}}$), combined with a smaller projected separation $(<3~\mathrm{kpc})$, may imply that they are two bright emission regions of a single galaxy. However, if components A and B were a single galaxy, the combined projected diameter would then be $\sim 11$kpc, which, as was the case for GRB~050820A, is far larger than what is typically observed at $z>2$. Also, given the morphology and dynamics indicated in Fig.~\ref{fig:posmaps}, it seems more likely that the three components identified in the data cube correspond to three closely interacting galaxies.

Similar to GRB 050820A, components A and B, which are unlikely to have hosted the GRB, are blueshifted from the likely host galaxy (component C). To construct a 3D picture of the galaxy system, we again compare the kinematics measured from our IFU data to those derived from the GRB afterglow spectrum presented in \cite{bolmer2019_150403abs_Z}. Absorption features are detected in the absorption spectrum at -60 km~s$^{-1}$ and -100 km~s$^{-1}$ from the GRB in several lines (e.g. \ion{O}{i}, \ion{Mn}{ii} and \ion{Mg}{ii}), consistent with the velocity offsets that we measure between the likely GRB host galaxy and components A and B, respectively. This would imply that components A and B lie in front of the host galaxy (component C), and thus as was the case for GRB~050820A, the host galaxy of GRB~150403A appears to be moving away from the other two companion galaxies. Absorption features at higher velocities at -150 and -200 km~s$^{-1}$ are also observed relative to component C. However, these are undetected in emission and thus correspond to diffuse gas that may originate from galaxy inflows/outflows, possibly related to interactions between components A and B. 

\subsubsection{Examples of interacting GRB host galaxies}
There are examples of other GRB hosts that show evidence of interaction with another galaxy from their morphologies or kinematics, such as GRB 980425\citep{arabsalmani2015_980425interact, arabsalmani2019_980425interact}, GRB 980613 \citep{djorgovski2003_merger1}, GRB 990123 \citep{bloom1999merger2, fruchter1999merger3}, GRB 080810 \citep{wiseman2017_interacting080810}, GRB100219A \citep{thone2013_interacting100219} and GRB 171205A \citep{arabsalmani2022_171205Ainteract}. Similar to the host galaxies studied in this paper, GRB 980425, GRB 980613, GRB 990123 and GRB 080810 were also found in host galaxies that are surrounded by multiple neighbouring components. Additionally, afterglow absorption spectrum of GRB 090323 and GRB 080810 showed absorption features at large velocity offsets indicative of a companion galaxy \citep{savaglio2012supersolar_interacting, wiseman2017_interacting080810}.

The hosts of GRB 050820A and GRB 150403A show evidence of interactions, both through the morphologies of their observed systems and through the absorbing features seen in the GRB afterglow spectra. Merging or interacting galaxies can trigger star formation \citep{somerville2001_interaction+starburst, teyssier2010_int+starburst+sim}, and it may therefore be reasonable to expect an increase in long GRBs occurring within interacting galaxy systems. Furthermore, given the increasing merger rate with redshift up to $z \sim 2$ \citep{le2000_mergerrate}, it is likely that more interacting galaxies will be observed as GRB hosts in the high-redshift ($z \sim 2-3$) universe. Quantifying the importance of galaxy interactions in the formation of long GRBs will require a complete study on the morphological and kinematic properties of GRB host galaxies relative to the general star forming galaxy population.

\subsection{Implications for the \texorpdfstring{$0.3Z_{\odot}$}{0.3Zsun} theoretical threshold}
The collapsar model predicts a $0.3Z_{\odot}$ metallicity threshold for progenitors of long GRBs \citep{woosley1993_GRB+collapsarmodel}. Consequently, a similar cut-off is expected in the star-forming regions hosting a GRB event. For GRB~050820A, both the emission and absorption metallicity measurements lie close to $0.3Z_{\odot}$ ($\sim 8.21$). However, the majority of the spaxels fall below this value (see top right panel in Fig.~\ref{fig:R23_panel}), and for component C, the measured metallicity is consistent with being below $0.3Z_{\odot}$ when considering uncertainties. This therefore suggests that our results for GRB~050820A are not in strong tension with the standard collapsar model.

For GRB~150403A, the metallicity of component C is not well constrained and we therefore report this as an upper limit. The remaining two components appear to have metallicities above the collapsar model theoretical threshold (see the bottom histograms in Figs.~\ref{fig:R23_panel} and \ref{fig:Rhat_panel}), although consistent within the uncertainties (see Table~\ref{tab:Metal_SFR}). They are, however, significantly larger than the absorption line metallicity, which may imply that these two components are more metal-rich than component C.

Although the metallicities of both GRB hosts lie close to or above the $0.3Z_{\odot}$ threshold, several previous observations of GRB hosts, both nearby ($z<1$) and more distant ($z>2$), have reported metallicities comparable to or higher than $0.3Z_{\odot}$ \citep[e.g,][]{kruhler2015grb}. The discrepancies between emission and absorption line metallicities introduce significant uncertainty regarding whether the emission line metallicities truly trace the metallicity of the progenitor environment. Considering these factors, along with the uncertainties in high-redshift metallicity diagnostics, we do not consider the metallicities measured in this paper to be significantly unusual.

\subsection{The characteristic host galaxy properties}
The flux estimates trace the SFR as the emission is most likely due to ionization by young stars. For the observations of GRB 050820A, the component within which the GRB is located is the faintest region among all three components. However, this does not necessarily indicate that the GRB happened in a low star-forming region consistent with many previous observations \citep[e.g.,][]{bloom+2002_LGRB_starformation, fruchter2006_GRB_massive, kruhler2015grb, kruhler2017_museGRB}. The estimates done using the H$\alpha$ flux yield a SFR of $5.4 \pm 2.0 ~\mathrm{M}_{\odot}\mathrm{yr}^{-1}$ for the component associated with the GRB location. Other hosts have been observed in the past for which the observed location of the GRB was found not to lie within the region with the highest SFR \citep[e.g.][]{kruhler2011_GRB_SFR,izzo2017_museGRB}. Nevertheless, component C is still a region with significant star formation. Compared to the larger population of long GRB hosts studied by \cite{kruhler2015grb}, the SFR of component C appears lower than the median value of $\sim 20 ~\mathrm{M}_{\odot}\mathrm{yr}^{-1}$ observed for long GRB hosts at $z > 2$. This lower SFR could be attributed to the host galaxy of GRB 050820A being a tidal dwarf galaxy, as suggested by \cite{chen2012_GRB050820A_HST}, since typical long GRB hosts generally exhibit higher SFRs than tidal dwarf galaxies \citep{ploeckinger2015TDG_SFR_sim, lee2016TDG_SFR_obs}. Alternatively, the host galaxy may still be in the process of transitioning to a higher SFR due to the ongoing interactions with neighbouring galaxies, similar to the starburst phase transitions observed in particularly massive GRB hosts by \cite{nadolny2023_transitionSB}.

Component C also has a mass estimate of $\log(M_\star/M_\odot)\sim 9.3$ from the literature \citep{chen2009_GRB050820}. Using this estimate and the work done by \cite{thorne2021deep_DEVILS} from the Deep Extragalactic Visible Legacy Survey (DEVILS), it is possible to check where this host lies on the galaxy main sequence. We see that the SFR estimated in this work lies slightly below the point where the number density of galaxies peaks in galaxy main sequence plots at the redshift bins of both $2.2 < z < 2.6$ and $2.6 < z < 3.25$ \citep[see Fig.~13 in][]{thorne2021deep_DEVILS}. Similarly, comparing with $z \sim 2.3$ galaxies from the MOSDEF survey with a mass range $\log(M_\star/M_\odot)$ = 9.15 -- 9.68 also reveals that component C has a lower SFR than the measured median value of 11.6 $\mathrm{M}_{\odot}\mathrm{yr}^{-1}$ \citep{sanders2015mosdef}.

On the other hand, had the galaxy complex not been imaged with {\em HST} beforehand, revealing multiple extended components, the {\em JWST} follow-up may well have been planned in slit-based rather than IFU mode. In such a scenario, component B might have been incorrectly assigned as the host galaxy instead of component C, and the resulting SFR would be consistent with both the MOSDEF and the DEVILS samples as well as other long GRB hosts within \cite{kruhler2015grb}. Even with the {\em HST} observation, a fixed slit observation of the host galaxy could have included contamination from component B, leading to an overestimation of the measured SFR. This may already be the case for some high redshift ($z >2$) GRB host studies that lack spatially resolved observations. If the SFR in these hosts is overestimated due to contamination, it would mean that the SFR of component C is closer to being consistent with other high-redshift ($z > 2$) GRB hosts. 

Observations of GRB 150403A show that component C, with which the GRB is affiliated, shows significant line emission and a SFR which is $39^{+76}_{-36} ~\mathrm{M}_{\odot}\mathrm{yr}^{-1}$. The SFR is largely unconstrained due to the uncertainty on host galaxy dust corrections. However, the uncertainty at the lower end is set by the fact that the SFR cannot be lower than the dust uncorrected H$\alpha$ flux, corresponding to a SFR $\sim 3 ~\mathrm{M}_{\odot}\mathrm{yr}^{-1}$. This shows that GRB 150403A is also affiliated with a star-forming region, consistent with GRB 050820A and previous long GRB host galaxy observations \citep[e.g.,][]{sollerman2005_lowZ, gorosabel2005_lowZ, kruhler2011_GRB_SFR}. Currently, no mass estimates are present in the literature for the host of GRB 150403A, by lack of sensitive long-wavelength photometric measurements of the host.

As was the case for GRB 050820A, if the afterglow of GRB 150403A had not been well localized, it would be unclear which of the three galaxies was the host. Given the separation of component C from the other two components, contamination from neighbouring regions is not expected. However, misidentifying the host galaxy would have been a possibility. 

\section{Conclusions}\label{sec:conclusion}
We present our spectroscopic analysis of the intermediate redshift host galaxies of GRB 050820A and GRB 150403A observed with {\em JWST}/NIRSpec using IFU data. These IFU observations are among the first observations where we were able to resolve multiple star-forming components at the GRB location within redshift $z >2$ galaxies instead of single slit observations. This provided the opportunity of comparing the ISM properties and characteristics such as metallicities of these resolved regions with single slit observations of high redshift GRB hosts from earlier works. In addition, we also compared the line diagnostic properties of our sample with larger samples characterizing the broader star-forming populations in the local Universe and at $z \sim 2.3$. Having IFU observations further allowed for the kinematics of the observed systems to be studied making it possible to build a 3D picture of the interacting galaxies.

Based on our analysis of the velocity offsets and the spatial separation between the multiple star-forming components in each observation, we believe that the observed targets are interacting with one another and that it is likely that all components are independent galaxies. The exception to this could be the host of GRB 050820A which appears to be a tidal dwarf galaxy that formed during the interactions of the other two galaxies, which is also suggested in \cite{chen2012_GRB050820A_HST}. Comparisons between the spatial and velocity offsets of the galaxies measured in our IFU data combined with the GRB afterglow absorption kinematics suggest that, in both cases, the neighbouring galaxies may lie between us and the host galaxies. Additionally, both GRB hosts appear to be moving away from their neighbouring galaxies.

Both GRB hosts were observed to be found in star-forming galaxies in line with expectations and previous observations \citep{le2003_GRBhostSFR, fruchter2006_GRB_massive, perley2016_GRB_SFR}. Furthermore, the ISM properties of the GRB host galaxies are consistent with other GRB host galaxies and star-forming galaxies found at $z > 2$. 

For GRB 050820A, we find that both the metallicity at the GRB location and the metallicity of the overall host galaxy is consistent with the absorption metallicity of $8.20 \pm 0.10$ \citep{wiseman2017abs_metallicity}. For GRB 150403A, we report the metallicity of the host galaxy as upper limits due to non-detection of H$\beta$. However, we note that the neighbouring galaxies (components A and B) appear to have metallicities that are $\sim 0.6$~dex higher than the absorption metallicity of $7.77 \pm 0.05$ \citep{bolmer2019_150403abs_Z}, which may imply that they are more-metal rich than the GRB host galaxy (i.e. component C).

In some of our measurements of the SFRs and certain line diagnostics, we observed differences between the values obtained for individual components and those averaged across all components. This shows the importance of spatially resolved spectroscopy, particularly for high redshift objects. IFU observations allow to create maps of parameters such as metallicity, SFRs, and line ratios, revealing spatial variations across a galaxy. Furthermore, the kinematic information content provides insights into dynamical characteristics, such as rotational patterns, velocity dispersions, and evidence of interactions or mergers. This information can improve our understanding of high redshift GRB host galaxies and of the broader galaxy population traced by long GRBs. As observational campaigns continue to expand with instruments like {\em JWST}, we may find more GRB host galaxies near star-forming companions like those observed in this study. This could provide further evidence supporting the hypothesis that GRB hosts are often irregular galaxies \citep{conselice_2005morphologies, fruchter2006_GRB_massive, wainwright_2007morphologies}, possibly shaped by past interactions with neighbouring galaxies. Therefore, it is essential to conduct such spectroscopic analyses with larger samples, which would also allow us to obtain more robust conclusions about how high redshift GRB hosts differ from their local counterparts or how they differ from broader galaxy populations at similar redshifts.


\section*{Acknowledgements}
PS acknowledges support from the UK Science and Technology Facilities Council, grant reference ST/X001067/1. AR acknowledges support by PRIN-MIUR 2017 (grant 20179ZF5KS). RLCS is supported by the Leverhulme Trust, grant RPG-2023-240. LC is supported by DFF/Independent Research Fund Denmark, grant-ID 2032–00071. The Cosmic Dawn Center is funded by the Danish
National Research Foundation under grant no. 
This work makes use of the Python packages {\sc numpy} \citep[v.1.26.4;][]{numpy}, {\sc matplotlib} \citep[v.3.8.0;][]{matplotlib}, {\sc uncertainties} (v.3.1.7; \url{http://pythonhosted.org/uncertainties}), {\sc specutils} \citep[v.1.13.0][]{specutils}, {\sc pandas} \citep[v.2.2.1][]{reback2020pandas} and {\sc astropy} \citep[v.5.3.4][]{astropy:2013, astropy:2018, astropy:2022}.

\section*{Data Availability}
The data underlying this article are available in the MAST Data Discovery Portal at https://stdatu.stsci.edu/datadownloads.html, and can be accessed with proposal ID 2344.

\bibliographystyle{mnras}
\bibliography{Bibliography.bib}

\appendix
\section{\textbf{GAUSSIAN FITS TO STACKED SPECTRA}}
\begin{figure*}
    \centering
    \centering
    \includegraphics[width=0.9\textwidth]{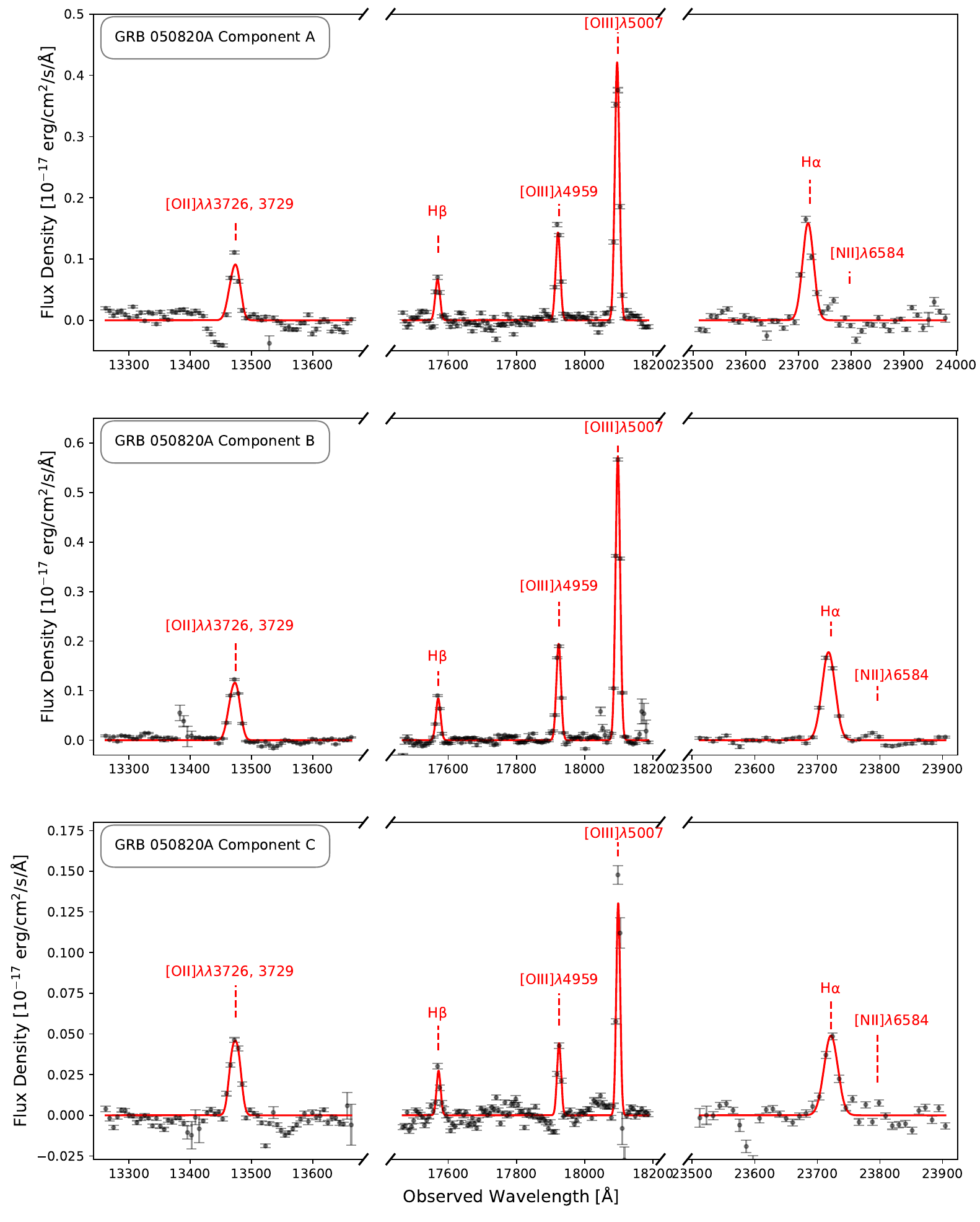}
    \caption{Stacked spectra of all three components of GRB 050820A. The red curves show Gaussian fits to detected emission lines, ranging from [\ion{O}{ii}]$\lambda\lambda3726, 3729$ to H$\alpha$. The expected position of the undetected [\ion{N}{ii}]$\lambda6584$ line is also marked in each spectrum. In both cases, the best-fit background has been subtracted, centering the noise level at zero. \textit{Note}: For component C, the best fit amplitude of [\ion{O}{iii}]$\lambda5007$ appears slightly lower than the top data point due to lower limit of its width being constrained by the instrument LSF.}
    \label{fig:1dspec_05}
\end{figure*}

\begin{figure*}
    \centering
    \centering
    \includegraphics[width=0.9\textwidth]{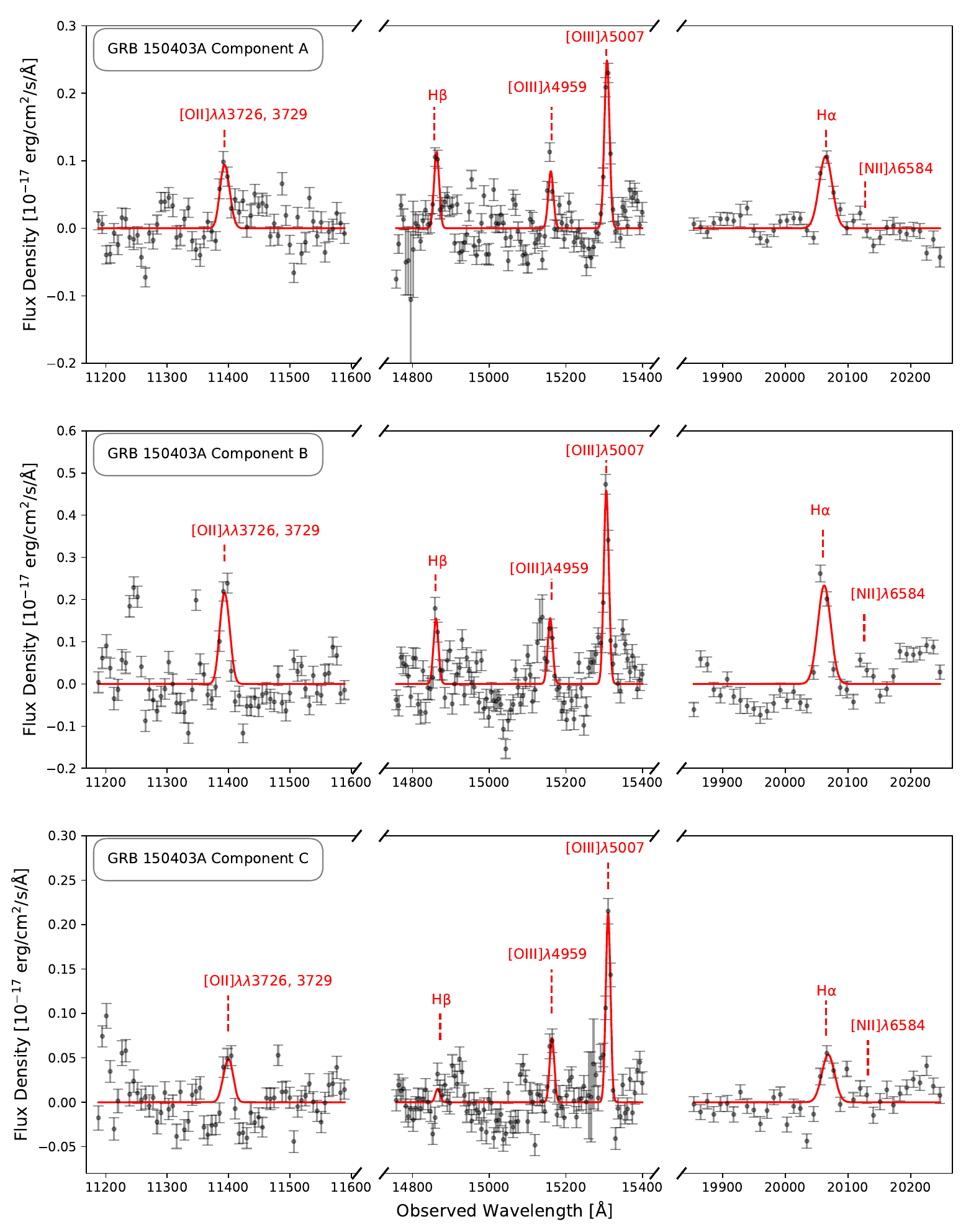}
    \caption{Same as Fig.~\ref{fig:1dspec_05} but for GRB 150403A.}
    \label{fig:1dspec_15}
\end{figure*}

\section{LOWER BRANCH METALLICITIES FROM THE SST24 DIAGNOSTIC}
\begin{table*}
    \centering
    \caption{Table with lower-branch SST24 R$_{23}$ \citep{sanders2024SST23} emission-line metallicities along with the corresponding $\mathrm{R}_{23}$ line ratios that we measure for the total integrated and individual components detected in the host galaxies of GRB 050820A and GRB 150403A.}
    \label{tab:SST24_lower}
    \resizebox{0.4\textwidth}{!}{
    \begin{tabular}{lccccccc}
    \hline
    \multicolumn{1}{c}{\multirow{2}{*}{GRB Host}} & \multirow{2}{*}{R$_{23}$} & 
    \multicolumn{1}{c}{12+log(O/H)}   \\ \cline{3-3}
     \multicolumn{1}{c}{} & \multicolumn{1}{c}{} & \multicolumn{1}{c}{SST24 R$_{23}$}  \\ \hline 
     050820A & &  \\ \hspace{3mm}total integrated & $1.04 \pm 0.07$ & $7.91 \pm 0.19$  \\
    \hspace{3mm}component A & $1.02 \pm 0.12$ & $7.79 \pm 0.28$ \\
    \hspace{3mm}component B & $1.02 \pm 0.10$ & $7.82 \pm 0.27$ \\
    \hspace{3mm}component C & $0.96 \pm 0.27$ & $7.58 \pm 0.45$ \\
    150403A & & &\\ \hspace{3mm}total integrated & $0.96 \pm 0.87$ & $7.35 \pm 0.78$ \\ 
    \hspace{3mm}component A & $0.62 \pm 0.09$ & $6.95 \pm 0.13$ \\
    \hspace{3mm}component B & $0.80 \pm 0.11$ & $7.26 \pm 0.24$ \\
    \hspace{3mm}component C & $1.77 \pm 1.71$ & $< 8.35$ \\ 
    \hline 
    \end{tabular}}
\end{table*}

\bsp	
\label{lastpage}
\end{document}